# Distributed Deterministic Edge Coloring using Bounded Neighborhood Independence

Leonid Barenboim[*]    Michael Elkin[*]

November 8, 2018

## Abstract

We study the *edge-coloring* problem in the message-passing model of distributed computing. This is one of the most fundamental and well-studied problems in this area. Currently, the best-known deterministic algorithms for $(2\Delta - 1)$-edge-coloring requires $O(\Delta) + \log^* n$ time [24], where $\Delta$ is the maximum degree of the input graph. Also, recent results of [5] for vertex-coloring imply that one can get an $O(\Delta)$-edge-coloring in $O(\Delta^\epsilon \cdot \log n)$ time, and an $O(\Delta^{1+\epsilon})$-edge-coloring in $O(\log \Delta \log n)$ time, for an arbitrarily small constant $\epsilon > 0$.

In this paper we devise a drastically faster deterministic edge-coloring algorithm. Specifically, our algorithm computes an $O(\Delta)$-edge-coloring in $O(\Delta^\epsilon) + \log^* n$ time, and an $O(\Delta^{1+\epsilon})$-edge-coloring in $O(\log \Delta) + \log^* n$ time. This result improves the state-of-the-art running time for deterministic edge-coloring with this number of colors in almost the entire range of maximum degree $\Delta$. Moreover, it improves it *exponentially* in a wide range of $\Delta$, specifically, for $2^{\Omega(\log^* n)} \leq \Delta \leq polylog(n)$. In addition, for small values of $\Delta$ (up to $\log^{1-\delta} n$, for some fixed $\delta > 0$) our deterministic algorithm outperforms all the existing *randomized* algorithms for this problem.

On our way to these results we study the *vertex-coloring* problem on the family of graphs with bounded *neighborhood independence*. This is a large family, which strictly includes line graphs of $r$-hypergraphs (i.e., hypergraphs in which each hyperedge contains $r$ or less vertices) for any $r = O(1)$, and graphs of bounded growth. We devise a very fast deterministic algorithm for vertex-coloring graphs with bounded neighborhood independence. This algorithm directly gives rise to our edge-coloring algorithms, which apply to *general* graphs.

Our main technical contribution is a subroutine that computes an $O(\Delta/p)$-defective $p$-vertex coloring of graphs with bounded neighborhood independence in $O(p^2) + \log^* n$ time, for a parameter $p$, $1 \leq p \leq \Delta$. In all previous efficient distributed routines for $m$-defective $p$-coloring the product $m \cdot p$ is super-linear in $\Delta$. In our routine this product is *linear* in $\Delta$, and this enables us to speed up the coloring drastically.

[*]Department of Computer Science, Ben-Gurion University of the Negev, POB 653, Beer-Sheva 84105, Israel. E-mail: {leonidba,elkinm}@cs.bgu.ac.il
This research has been supported by the Binational Science Foundation, grant No. 2008390.

# 1 Introduction

## 1.1 Edge-Coloring

We study the *edge-coloring* problem in the *message passing model* of distributed computing. Specifically, we are given an $n$ vertex undirected unweighted graph $G = (V, E)$, with each vertex hosting an autonomous processor. The processors have distinct identity numbers (henceforth, Ids) from the range $\{1, 2, ..., n\}$. They communicate with each other over the edges of $E$. The communication occurs in discrete rounds. In each round each vertex can send a message to each of its neighbors, and these messages arrive to their destinations before the next round starts. The running time of an algorithm in this model is the number of rounds of communication that are required for the algorithm to terminate.

A legal *edge-coloring* $\varphi$ of $G = (V, E)$ is a function $\varphi : E \to N$ that satisfies that for any pair of edges $e, e' \in E$ that share an endpoint (henceforth, *incident*), it holds that $\varphi(e) \neq \varphi(e')$. Denote by $\Delta = \Delta(G)$ the maximum degree of the graph $G$. A classical theorem of Vizing [30] shows that for any graph $G$, its edges can be legally colored in $(\Delta + 1)$ colors. Obviously, at least $\Delta$ colors are required.

The edge coloring problem is one of the most fundamental problems in Graph Theory and Graph Algorithms. It also has numerous applications in Computer Science, including job-shop scheduling, packet-routing, and resource allocation [16, 11]. This problem was also extensively studied in the message-passing model. Panconesi and Rizzi [24] showed that a $(2\Delta - 1)$-edge-coloring can be computed deterministically in $O(\Delta) + \log^* n$ time. Panconesi and Srinivasan [26] devised a randomized $(1.6\Delta + O(\log^{1+\epsilon} n))$-edge coloring algorithm that runs in polylogarithmic time, where $\epsilon > 0$ is an arbitrarily small constant. Dubhashi et. al. [10] used the Rődl nibble method to improve this to a randomized $(1 + \epsilon)\Delta$-edge-coloring in time $O(\log n)$, as long as $\Delta = \omega(\log n)$. Grable and Panconesi [14] showed that if for every edge $e = (u, w)$, the degree of either $u$ or $w$ is sufficiently large (at least $2^{\Omega(\frac{\log n}{\log \log n})}$), then $(1 + \epsilon)\Delta$-edge-coloring, for an arbitrarily small constant $\epsilon > 0$, can be computed in $O(\log \log n)$ time by a randomized algorithm. Czygrinow et. al. [9] devised a deterministic $O(\Delta \log n)$-edge-coloring that requires $O(\log^4 n)$ time.

A more general approach to the edge-coloring and many other related problems was taken in [1, 22, 25]. These papers presented algorithms that compute a *network decomposition*, i.e., a partition of the input graph into regions of small diameter. This partition admits also additional helpful properties. This partition can then be used to compute edge-coloring, vertex-coloring, maximal independent set, and other related structures. In particular, by this technique one can get a deterministic $(2\Delta - 1)$-edge-coloring algorithm that requires $2^{O(\sqrt{\log n})}$ time [25, 1].

A (legal) vertex coloring $\psi$ of $G = (V, E)$ is a function $\psi : V \to N$ that satisfies that for any edge $e = (u, w) \in E$, $\psi(u) \neq \psi(w)$. We refer to $\psi(u)$ as the $\psi$-*color of* $u$. By considering the line graph $L(G) = (E, \mathcal{E} = \{(e, e') \mid e \cap e' \neq \emptyset\})$, it is easy to see that any vertex-coloring algorithm that employs $f(\Delta)$ colors, for a function $f()$, translates into an edge-coloring algorithm that employs $f(2\Delta)$ colors, with essentially the same running time [1]. This observation enables one to harness many of the recent advances in vertex-coloring for obtaining significantly faster edge-coloring algorithms as well. Most relevant in this context are the results of [18, 29, 5]. Kothapalli et. al. [18] showed that an $O(\Delta)$-vertex-coloring (and, consequently, $O(\Delta)$-edge-coloring as well) can be computed in $O(\sqrt{\log n})$ rounds, by a randomized algorithm. Recently Schneider and Wattenhofer [29] devised a randomized algorithm that computes (1) a $(\Delta+1)$-vertex-(and edge-) coloring in $O(\log \Delta + \sqrt{\log n})$ time; (2) an $O(\Delta + \log n)$-coloring in $O(\log \log n)$ time; and (3) $O(\Delta \log^{(k)} n + \log^{1+1/k} n)$-coloring in $f(k) = O(1)$ time, for some fixed function $f()$ and any positive integer $k$. In [5] the authors of the current paper devised a deterministic algorithm that, for an arbitrarily small constant $\epsilon > 0$, computes (1) an $O(\Delta^{1+\epsilon})$-coloring in $O(\log \Delta \log n)$ time; and (2) an $O(\Delta)$-coloring in $\Delta^\epsilon \log n$ time.

In the current paper we show that in the case of *edge*-coloring the factor $\log n$ can be eliminated. Specifically, we devise a deterministic algorithm that for an arbitrarily small constant $\epsilon > 0$, computes

---
[1] As long as one allows arbitrarily large messages.

| Range of $\Delta$ | $\omega(\log^* n) = \Delta = o(\log n \log \log n)$ | $\Omega(\log n \log \log n) = \Delta$ |
|---|---|---|
| Previous | $(2\Delta - 1)$ colors, $O(\Delta) + \log^* n$ time [24] | $O(\Delta)$ colors, $O(\Delta^\epsilon \log n)$-time [5] |
| | | $O(\Delta^{1+\epsilon})$ colors, $O(\log \Delta \log n)$-time [5] |
| **New** | $O(\Delta)$ colors, $O(\Delta^\epsilon) + \log^* n$ time | $O(\Delta)$ colors, $O(\Delta^\epsilon) + \log^* n$ time |
| | $O(\Delta^{1+\epsilon})$ colors, $O(\log \Delta) + \log^* n$ time | $O(\Delta^{1+\epsilon})$ colors, $O(\log \Delta) + \log^* n$ time |

Table 1: *A concise comparison of previous state-of-the-art edge-coloring* deterministic *algorithms with our new algorithms.*

(1) an $O(\Delta^{1+\epsilon})$-edge-coloring in $O(\log \Delta) + \log^* n$ time; (2) an $O(\Delta)$-edge-coloring in $O(\Delta^\epsilon) + \log^* n$ time. In addition we have a tradeoff curve with a number of results along it, in whcih the number of colors is larger than $\Omega(\Delta)$, but smaller than $\Delta^{1+\epsilon}$.

These results compare very favorably to the state-of-the-art. We start with comparing them to deterministic algorithms. For $\Delta$ in the range $\omega(\log^* n) \leq \Delta \leq O(\log n \log \log n)$ the fastest currently known algorithm for edge-coloring with $O(\Delta^{1+\epsilon})$ or less colors is due to Panconesi and Rizzi [24]. Its running time is $O(\Delta) + \log^* n$. Our algorithm runs *exponentially faster*, in time $O(\log \Delta) + \log^* n$, but it employs more colors ($O(\Delta^{1+\epsilon})$ instead of $(2\Delta - 1)$). In addition, another variant of our algorithm employs only $O(\Delta)$ colors, and has a significantly better running time than that of the algorithm of [24], specifically, $O(\Delta^\epsilon) + \log^* n$. For $\Delta$ in the range $\Delta = \Omega(\log n \log \log n)$ the fastest known algorithm for edge-coloring with $O(\Delta^{1+\epsilon})$ colors is due to [5]. Its running time is $O(\log \Delta \log n)$, instead of $O(\log \Delta) + \log^* n$ for our new algorithm. Note that as long as $\Delta$ is at most polylogarithmic in $n$, the new running time is $O(\log \log n)$ instead of $O(\log n \log \log n)$ of [5], i.e., our improvement in this range is exponential as well. To summarize, our algorithm improves the state-of-the-art running time for deterministic algorithms in almost the entire range of the maximum degree $\Delta$, i.e., for $\Delta = \omega(\log^* n)$, and it impoves it exponentially for $2^{\Omega(\log^* n)} \leq \Delta \leq O(\log^k n)$, for an arbitrarily large constant $k$. See Table 1 for a concise comparison of previous and new deterministic results.

Next, we compare the running time and the number of colors of our *deterministic* algorithm with the state-of-the-art with respect to *randomized* algorithms. For $\Delta = \Omega(\log n)$ the recent randomized algorithm of Schneider and Wattenhofer [29] outperforms our algorithm. However, for $\Delta \leq \log^{1-\delta} n$, for an arbitrarily small constant $\delta > 0$, the algorithm of [29] either employs $\Omega(\log n)$ colors (i.e., more than $\Delta^{1+\epsilon}$ for an arbitrarily small $\epsilon > 0$), or its running time is $\Omega(\sqrt{\log n})$. (Note, however, that the randomized algorithm of [29] solves a generally harder vertex-coloring problem, rather than edge-coloring.) Hence in the range $\omega(\log^* n) \leq \Delta \leq \log^{1-\delta} n$, for some fixed constant $\delta > 0$, our deterministic algorithm outperforms all previous algorithms, deterministic and randomized. Moreover, in the range $2^{\Omega(\log^* n)} \leq \Delta \leq \log^{1-\delta} n$ our algorithm is *exponentially faster* than the previous ones. Indeed, for $\Delta \leq \sqrt{\log n}$ the best previous algorithm that achieves $O(\Delta^{1+\epsilon})$ or less colors is due to [24], whose running time is $O(\Delta) + \log^* n$. On the other hand, the running time of our algorithm is $O(\log \Delta) + \log^* n$. For $\sqrt{\log n} \leq \Delta \leq \log^{1-\delta} n$ the best previous algorithms that achieve that many colors are due to [29, 18], and their running time is $O(\sqrt{\log n})$. Our algorithm requires in this range just $O(\log \Delta) + \log^* n = O(\log \log n)$ time. (On the other hand, the variant of the algorithm of [29] that runs in $O(\sqrt{\log n})$ time employs just $(2\Delta - 1)$ colors, as opposed to $\Delta^{1+\epsilon}$ colors that are employed by our algorithm. The algorithm of [24] also employs only $(2\Delta - 1)$ colors.) See Table 2 for a concise comparison.

We remark also that all our aformentioned algorithms employ messages of size $O(\log n)$. On the other hand, converting an algorithm for vertex-coloring into an edge-coloring algorithm incurs an overhead of a factor $\Delta$ in the message size. In other words, if one adapts naively the algorithms of [29], [5], or [18] to the edge-coloring problem, he will get an algorithm that sends messages of size $O(\Delta \log n)$. Also, some of the state-of-the-art algorithms for edge-coloring that we mentioned employ large messages. In particular, this is the case for [9]. The general algorithm of Panconesi and Srinivasan [25] that computes a network decomposition and employs it to build a vertex- or an edge-coloring, also employs large messages (on the stage that uses decomposition to compute a coloring).

| Range of $\Delta$ | $\omega(\log^* n) = \Delta = O(\sqrt{\log n})$ | $\Omega(\sqrt{\log n}) = \Delta \leq \log^{1-\delta} n$ |
|---|---|---|
| Previous | $(2\Delta - 1)$ colors, $O(\Delta) + \log^* n$ time [24] | $(2\Delta - 1)$ colors, $O(\sqrt{\log n})$ time [29] |
| **New (Deter.)** | $O(\Delta^{1+\epsilon})$ colors, $O(\log \Delta) + \log^* n$ time | $O(\Delta^{1+\epsilon})$ colors, $O(\log \log n)$ time |

Table 2: *A concise comparison of previous state-of-the-art edge-coloring randomized and deterministic algorithms with our new deterministic algorithm. The algorithm of [24] is deterministic. The algorithm of [29] is randomized.*

Observe also that the $\log^* n$ term in the running time of our algorithms is optimal up to a factor of 2, in view of the lower bounds of [21]. Specifically, Linial's lower bound [21] implies that $f(\Delta)$-edge-coloring, for any fixed function $f()$, requires at least $\frac{1}{2} \log^* n$ time. Moreover, there is a variant of our algorithm that achieves the precisely optimal additive term of $\frac{1}{2} \log^* n$, while achieving almost the same dependence on $\Delta$. This variant, however, employs messages of size $O(\Delta \log n)$. By "almost" the same dependence on $\Delta$ we mean that it achieves (for an arbitrarily small constant $\epsilon > 0$), (1) an $O(\Delta)$-edge-coloring in $O(\Delta^\epsilon) + \frac{1}{2} \log^* n$ time, and (2) an $O(\Delta^{1+\epsilon})$-edge-coloring in $O(\log \Delta \cdot \frac{\log^* \Delta}{\log(\log^* \Delta)}) + \frac{1}{2} \log^* n$ time. In other words, item (1) is the same as that cited above, except that $\log^* n$ is replaced by $\frac{1}{2} \log^* n$, and in item (2) there is also a tiny slack factor of $\frac{\log^* \Delta}{\log(\log^* \Delta)}$.

### 1.2 Graphs with Bounded Neighborhood Independence

Our results for edge-coloring follow from far more general results that we describe below. *Neighborhood independence* $I(G)$ of a graph $G = (V, E)$ is the maximum number of independent [1] neighbors of a single vertex $v \in V$. The family of graphs with constant neighborhood independence (henceforth, *bounded neighborhood independence*) is a very general family of graphs. Indeed, for any graph $G$, the neighborhood independence of its line graph $L(G)$ is at most 2. Moreover, for an $r$-hypergraph $\mathcal{H}$ (i.e., a hypergraph in which every hyperedge contains at most $r$ vertices), $I(L(\mathcal{H})) \leq r$.

Another important family of graphs which is subsumed by the family of graphs with bounded neighborhood independence is the family of graphs of *bounded growth*. A graph $G = (V, E)$ is said to be of bounded growth if there exists a function $f()$ such that for any $r = 1, 2, ...$, the number of independent vertices at distance at most $r$ from any given vertex is at most $f(r)$. Distributed algorithms for vertex-coloring and computing a maximal independent set on graphs from this family is a subject of intensive recent research [17, 13, 28]. The crowning result of this effort is the deterministic algorithm of [28] that computes a maximal independent set and a $(\Delta + 1)$-vertex-coloring for graphs from this family in optimal time $O(\log^* n)$. Note, however, that a graph $G$ with a constant neighborhood independence may contain an arbitrarily large independent set $U$ whose all vertices are at distance at most 2 from some given vertex $v$ in $G$. See Figure 1 below for an illustration. Thus, graphs with bounded neighborhood independence may have unbounded growth.

Yet another family of graphs which is subsumed by the family of graphs of bounded independence is the family of *claw-free* graphs. A graph is *claw-free* if it excludes $K_{1,3}$ as an induced subgraph. (In fact, for any $r = 2, 3, ...$, the family of graphs with independence at most $r$ is precisely the family of graphs that exclude induced $K_{1,r+1}$.) The family of claw-free graphs attracted enormous attention in Structural Graph Theory. See, e.g., the series of papers by Chudnovsky and Seymour, starting with [6]. It was also studied from algorithmic perspective. In particular, in [23] Minty devised a polynomial-time sequential algorithm for computing maximum independent set in claw-free graphs.

---

[1] Two vertices $u, w$ are *independent* in $G$ if $(u, w) \notin E$.

In this paper we devise a vertex-coloring algorithm for graphs of bounded neighborhood independence that computes (for an arbitrarily small constant $\epsilon > 0$) (1) an $O(\Delta)$-vertex-coloring in $O(\Delta^\epsilon) + \frac{1}{2}\log^* n$ time, and (2) an $O(\Delta^{1+\epsilon})$-vertex-coloring in $O(\log \Delta \cdot \frac{\log^* \Delta}{\log(\log^* \Delta)}) + \frac{1}{2}\log^* n$ time. Modulo some subtleties, these results imply our main results about edge-coloring described in Section 1.1. In addition, they apply to line graphs of $r$-hypergraphs for any constant $r$, to claw-free graphs, to graphs of bounded growth, and to many other graphs.

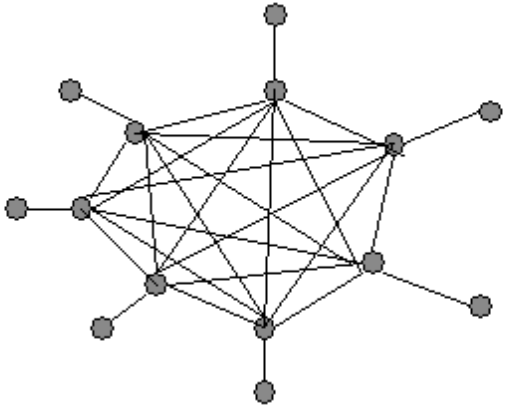

**Fig. 1.** A graph $G$ with $I(G) = c = 2$. The graph is obtained by connecting each vertex of an $n/2$-vertex clique with an isolated vertex. Each vertex in $G$ has at most 2 independent neighbors. However, each vertex $v$ in the clique has at least $n/2 = \Omega(\Delta)$ independent vertices in $\Gamma_2(v)$, and so the graph $G$ is not a graph of bounded growth.

### 1.3 Our Techniques

In the heart of our algorithms lie improved algorithms for computing *defective colorings*. For a non-negative integer $m$ and positive integer $\chi$, an *m-defective $\chi$-vertex-coloring* $\varphi$ of a graph $G = (V, E)$ is a function $\varphi : V \to \{1, 2, ..., \chi\}$ that satisfies that for every vertex $v \in V$, it has at most $m$ neighbors colored by $\varphi(v)$. The parameter $m$ is called the *defect* of the coloring. Defective coloring was introduced by Cowen et. al [7] and by Harary and Jones [15]. It was extensively studied from graph-theoretic perspective [2, 8, 12]. Recently defective coloring was discovered to be very useful in the context of distributed graph coloring [4, 19]. Specifically, the state-of-the-art $(\Delta+1)$-vertex-coloring algorithms for general graphs [4, 19] are based on subroutines for computing defective coloring. For a parameter $p$, these subroutines compute an $O(\Delta/p)$-defective $p^2$-coloring. (In [4] the running time of such a subroutine is $O(p^2) + \frac{1}{2}\log^* n$, and in [19] it is $O(\log^* \Delta) + \frac{1}{2}\log^* n$.) It was observed in [4] that one could have devised significantly faster coloring algorithms if there were an efficient (distributed) routine for computing a $m$-defective $\chi$-coloring with a *linear* in $\Delta$ product of $m$ and $\chi$. (The current state-of-the-art [4, 19], has $m = O(\Delta/p)$, $\chi = p^2$, i.e., $m \cdot \chi = O(\Delta \cdot p)$ instead of the desired $O(\Delta)$.)

In this paper we show that if one restricts his attention to the family of bounded neighborhood independence graphs, then this goal can be achieved. Specifically, we devise an algorithm that computes an $O(\Delta/p)$-defective $p$-vertex-coloring of a given graph of bounded neighborhood independence in $O(p^2) + \frac{1}{2}\log^* n$ time. As a result we obtain a bunch of drastically faster algorithms for vertex-coloring these graphs, and, consequently, for edge-coloring general graphs.

Whether it is possible to devise an efficient $O(\Delta/p)$-defective $p$-vertex-coloring algorithm for general graphs remains a challenging open question. Recently in [5] the authors of the current paper were able to circumvent this question by the means of *arbdefective coloring*. Note that an $O(\Delta/p)$-defective $p$-vertex-coloring can be seen as a partition of the vertex set into $p$ subsets, each inducing a subgraph of maximum degree at most $O(\Delta/p)$. In [5] the authors showed that the vertex set of a graph of arboricity [1] $a$ can be efficiently partitioned into $p$ subsets, each inducing a subgraph of arboricity $O(a/p)$. This partition is then employed in [5] to devise a suite of efficient algorithms for vertex-coloring general graphs. In particular, using this technique [5] devised an $O(\Delta^{1+\epsilon})$-vertex coloring algorithm in $O(\log \Delta \log n)$ time, for an arbitrarily small constant $\epsilon > 0$.

Note, however, that the factor of $\log n$ in the running time of the algorithms of [5] is inherent, because these algorithms rely heavily on the notion of arboricity, and more specifically, on the machinery of forest-decompositions developed in [3] for working with graphs of bounded arboricity. On the other hand, a lower bound shown in [3] stipulates that computing a forests-decomposition requires $\Omega(\frac{\log n}{\log a})$ time, where $a$ is the arboricity. Consequently, the factor of $\log n$ in the running time is unavoidable [1] using the approach of [5]. In the current paper we pursue a different line of attack. Specifically, we devise improved algorithms for *defective* coloring, rather than circumventing it and going through *arbdefective* coloring.

**1.4 Structure of the Paper**

In Section 2 we describe the definitions and notation employed in our algorithms. In Section 3 we present our defective vertex-coloring algorithms for graphs with bounded neighborhood independence. In Section 4 we devise legal vertex-coloring algorithms for this family of graphs. In Section 5 we devise legal edge-coloring algorithms for general graphs. In Section 6 we describe several extensions to our algorithms, including randomized variants.

## 2 Preliminaries

Unless the base value is specified, all logarithms in this paper are of base 2.

For a non-negative integer $i$, the *iterative log-function* $\log^{(i)}(\cdot)$ is defined as follows. For an integer $n > 0$, $\log^{(0)} n = n$, and $\log^{(i+1)} n = \log(\log^{(i)} n)$, for every $i = 0, 1, 2, ...$. Also, $\log^* n$ is defined by: $\log^* n = \min \left\{ i \mid \log^{(i)} n \leq 2 \right\}$.

The *degree* of a vertex $v$ in a graph $G = (V, E)$, denoted $deg(v) = deg_G(v)$, is the number of edges incident to $v$. A vertex $u$ such that $(u, v) \in E$ is called a *neighbor* of $v$ in $G$. The *neighborhood* $\Gamma(v) = \Gamma_G(v)$ of $v$ is the set of neighbors of $v$. The maximum degree of a vertex in $G$, denoted $\Delta(G)$, is defined by $\Delta = \Delta(G) = \max_{v \in V} deg(v)$. The graph $G' = (V', E')$ is a *subgraph* of $G = (V, E)$, denoted $G' \subseteq G$, if $V' \subseteq V$ and $E' \subseteq E$. The notation $V(G')$ and $E(G')$ is used to denote the vertex set $V'$ of $G'$, and the edge set $E'$ of $G'$, respectively.

The *line graph* $L(G) = (V'', E'')$ of a graph $G = (V, E)$ is a graph in which $V''$ contains a vertex $v_e$ for each edge $e \in E$, and an edge $(v_e, w_{e'})$ if and only if the edges $e$ and $e'$ of $E$ share a common endpoint. We say that a vertex $v_e \in V''$ and an edge $e \in E$ *correspond* to each other.

The *out-degree* of a vertex $v$ in a directed graph $\hat{G}$ is the number of edges incident to $v$ that are oriented outwards of $v$. An *orientation* $\sigma$ of (the edge set of) a graph is an assignment of direction to each edge $(u, v) \in E$, either towards $u$ or towards $v$. An edge $(u, v)$ that is oriented towards $v$ is denoted by $\langle u, v \rangle$. The *out-degree* of an orientation $\sigma$ of a graph $G$ is the maximum out-degree of a vertex in $G$ with respect to $\sigma$. In a given orientation, each neighbor $u$ of $v$ that is connected to $v$ by an edge oriented towards $u$ is called a *parent* of $v$. In this case we say that $v$ is a *child* of $u$.

---

[1] An *arboricity* of a graph $G = (V, E)$ is $a(G) = \max\{ \lceil \frac{|E(U)|}{|U|-1} \rceil : U \subseteq V, |U| \geq 2 \}$.

[1] In fact, the algorithm of [5] computes an $O(a^{1+\epsilon})$-coloring in time $O(\log a \log n)$ for graphs of arboricity $a$. An $O(\Delta^{1+\epsilon})$-coloring in $O(\log \Delta \log n)$ time is a direct corollary of this result. On the other hand, it is known [3] that $O(a^{1+\epsilon})$-coloring requires $\Omega(\frac{\log n}{\log a})$ time.

For a graph $G = (V, E)$, a set of vertices $U \subseteq V$ is called an *independent set* if for every pair of vertices $v, w \in U$ it holds that $(v, w) \notin E$.

The minimum number of colors that can be used in a legal vertex-coloring of a graph $G$ is called *the chromatic number* of $G$, denoted $\chi(G)$.

Next, we state a number of known results that will be used in our algorithms.

**Lemma 2.1.** *(1) [21] A legal $O(\Delta^2)$-vertex-coloring can be computed in $\log^* n$ time.*
*(2) [4, 19] A legal $(\Delta + 1)$-vertex-coloring can be computed in $O(\Delta) + \log^* n$ time.*
*(3) [19] A $\lfloor \Delta/p \rfloor$-defective $O(p^2)$-vertex-coloring can be computed in $O(\log^* n)$ time.*

## 3  Defective Coloring

In this section we present a defective vertex coloring algorithm for graphs with bounded neighborhood independence. We begin with a formal definition of this family of graphs.

**Definition 3.1. Graphs with neighborhood independence bounded by $c$.**
*For a graph $G = (V, E)$ and a vertex $v \in V$, the* neighborhood independence *of $v$, denoted $I(v)$, is the size of maximum-size independent subset $U \subseteq \Gamma(v)$ of neighbors of $v$.*
*The* neighborhood independence *of a graph $G$ is defined as $I(G) = \max_{v \in V} \{I(v)\}$. For a positive parameter $c$, a graph $G = (V, E)$ is said to have* neighborhood independence bounded by $c$ *if $I(G) \leq c$.*

Let $c$ be a fixed positive constant, and $p$ be a parameter such that $1 \leq p \leq \Delta$. We devise a procedure, called *Procedure Defective-Color*, that computes an $O(\Delta/p)$-defective $p$-coloring on graphs with neighborhood independence bounded by $c$. This coloring is achieved by first computing a defective $O(p^2)$-coloring, and then reducing the number of colors to $p$, using special properties of graphs with bounded neighborhood independence. Procedure Defective-Color receives as input a graph $G$ with neighborhood independence bounded by $c$, a positive parameter $b$, and the parameter $\Lambda$ which serves as an upper bound on the maximum degree of the input graph. The parameter $b$ satisfies that $b \geq 1$, $b \cdot p \leq \Lambda$. This parameter controls the tradeoff between the defect of the resulting coloring and the running time of the procedure. Specifically, the defect behaves as $\frac{\Lambda}{p}(1 + O(1/b))$, and the running time is at most $O(b^2 \cdot p^2 + \log^* n)$. We assume that all vertices know the value of $c$ before the computation starts.

The procedure starts with computing a $\lfloor \Lambda/(b \cdot p) \rfloor$-defective $O((b \cdot p)^2)$-coloring $\varphi$ of $G$ using Lemma 2.1(3). The coloring $\varphi$ is employed for computing another defective coloring $\psi$ of the vertices of $G$. The recoloring step spends one round for each $\varphi$-color class. Specifically, each vertex $v \in V$ computes $\psi(v)$ as follows. The vertex $v$ waits for each neighbor $u$ of $v$ with $\varphi(u) < \varphi(v)$ to select a color $\psi(u)$. Once $v$ receives a message from each such neighbor $u$ with its color $\psi(u)$, it sets $\psi(v)$ to be a value from the range $\{1, 2, ..., p\}$ that is used by the minimum number of neighbors $u$ with $\varphi(u) < \varphi(v)$. Once $v$ selects its color $\psi(v)$, it sends it to all its neighbors. This completes the description of the algorithm.

We need the following piece of notation. For a vertex $v$ and an index $k \in \{1, 2, ..., p\}$, let $N_v(k) = |\{u \in \Gamma(v) \mid \psi(u) = k, \varphi(u) < \varphi(v)\}|$ denote the number of neighbors $u$ of $v$ that have smaller $\varphi$-color than $v$ has, and whose $\psi$-color was set to $k$. Next, we provide the pseudocode of Procedure Defective-Color.

**Algorithm 1** Procedure Defective-Color$(G, b, p, \Lambda)$

An algorithm for each vertex $v \in V$.

1: $\varphi(v) :=$ compute $\lfloor \Lambda/(b \cdot p) \rfloor$-defective $O((b \cdot p)^2)$-coloring using Lemma 2.1(3)
2: send $\varphi(v)$ to all neighbors
3: $\psi(v) := 0$
4: **while** $\psi(v) = 0$, in each round **do**
5:    **if** $v$ received $\psi(u)$ for each neighbor $u$ of $v$ with $\varphi(u) < \varphi(v)$ **then**
6:       $m := \min\{N_v(k) \mid k \in \{1, 2, ..., p\}\}$
7:       $\psi(v) :=$ a color $k \in \{1, 2, ..., p\}$ such that $N_v(k) = m$
8:       send $\psi(v)$ to all neighbors
9:    **end if**
10: **end while**

Observe that a vertex $v$ waits only for neighbors with smaller $\varphi$-color before selecting $\psi(v)$. Consequently, it selects the color $\psi(v)$ after at most $\varphi(v)$ rounds from the time when step 2 of Algorithm 1 was executed. This fact is stated in the following lemma.

**Lemma 3.2.** *Let $\varphi$ be the coloring computed in the first step of Algorithm 1. Let $R$ be the round in which step 2 is executed. A vertex $v$ selects a color $\psi(v) \neq 0$ in round $R + \varphi(v)$ or earlier.*

*Proof.* Let $\ell = O((b \cdot p)^2)$ be the number of colors employed by $\varphi$. The lemma is proved by induction on the number of rounds. We prove that once a round $i = R+1, R+2, .., R+\ell$ is completed, all vertices $v$ with $\varphi(v) \leq i$ have already selected the color $\psi(v)$. For the base case, observe that the vertices $v$ with $\varphi(v) = 1$ have no neighbors with smaller $\varphi$-color. Therefore they select a $\psi$-color in round $R+1$, immediately after receiving the $\varphi$-colors of their neighbors in round $R$. For the induction step, suppose that in round $i-1$ all vertices $v$ with $\varphi(v) \leq i-1$ have already selected the color $\psi(v)$. Hence, each vertex $u$ with $\varphi(u) \leq i$ receives the color $\psi$ for each neighbor with smaller $\varphi$-color before round $i$. Consequently, the vertices $u$ select a $\psi$-color in round $i$ or earlier. □

Since the coloring $\varphi$ employs $\ell = O((b \cdot p)^2)$ colors, it follows that all vertices select a $\psi$-color at most $\ell$ rounds after the computation of the defective coloring in step 1 of Algorithm 1. By Lemma 2.1(3), step 1 requires $O(\log^* n)$ rounds. The overall running time of Algorithm 1 is given in the following corollary.

**Corollary 3.3.** *The running time of Procedure Defective-Color is $O((b \cdot p)^2 + \log^* n)$.*

In what follows we prove the correctness of Procedure Defective-Color. It is easy to see (using the Pigeonhole principle) that the number of neighbors $u$ of a vertex $v$ such that $\varphi(u) < \varphi(v)$ and $\psi(u) = \psi(v)$ is at most $\Lambda/p$. (Otherwise, there are at least $(\Lambda/p) + 1$ neighbors of $v$ that are colored by a $\psi$-color $i$, for each $i = 1, 2, ..., p$. Hence, $v$ has more than $\Lambda$ neighbors, a contradiction.) In addition, since $\varphi$ is a $\lfloor \Lambda/(b \cdot p) \rfloor$-defective coloring, there are at most $\lfloor \Lambda/(b \cdot p) \rfloor$ neighbors $u$ of $v$ that have the same $\varphi$-color as $v$ has, i.e., satisfy $\varphi(u) = \varphi(v)$. These neighbors may also end up selecting the same $\psi$-color that $v$ selects. Hence the number of neighbors $u$ of $v$ that satisfy $\varphi(u) \leq \varphi(v)$ and $\psi(u) = \psi(v)$ is at most $\Lambda/p + \Lambda/(b \cdot p)$. In order to prove that $\psi$ is an $O(\Lambda/p)$-defective coloring we also prove a somewhat surprising claim regarding the other neighbors of $v$. Specifically, we prove that the number of neighbors $u$ of a vertex $v$ such that $\varphi(u) > \varphi(v)$ and $\psi(u) = \psi(v)$ is $O(\Lambda/p)$ as well. Consequently, $\psi$ is an $O(\Lambda/p)$-defective $p$-coloring.

For $i = 1, 2, ..., p$, let $G_i$ be the subgraph induced by the $\psi$-color class $i$, i.e., by the vertex set $\{v \in V \mid \psi(v) = i\}$. As a first step we show that the chromatic number $\chi(G_i)$ of $G_i$ is at most $(\Lambda/(b \cdot p) + \Lambda/p) + 1$. We prove this claim by presenting an acyclic orientation of $G_i$ with out-degree at most $(\Lambda/(b \cdot p) + \Lambda/p)$. Since a graph with acyclic orientation with out-degree $d$ is legally $(d+1)$-colorable (see Lemma 3.4), the claim follows. Then we show that bounded chromatic number in conjunction with bounded neighborhood independence imply a bounded degree.

**Lemma 3.4.** *A graph $G$ with an acyclic orientation $\mu$ with out-degree $d$ satisfies that $\chi(G) \leq d+1$.*

*Proof.* We color the vertex set of $G$ as follows. A vertex $v$ waits for all its neighbors $u$ connected to $v$ by outgoing edges $\langle v, u \rangle$ to select a color. (See Figure 2 below.) Then it selects a color from the range $\{1, 2, ..., d+1\}$ that is not used by any such neighbor. (The number of neighbors $u$ as above is at most the out-degree of the orientation, that is at most $d$.) Since the orientation is acyclic, this process terminates and produces a $(d+1)$-coloring. The coloring is legal since for each edge $(u, v) \in E$, if it is oriented by $\mu$ towards $v$, the vertex $u$ selects a color that is different from the color of $v$. Otherwise, $v$ selects a color that is different from the color of $u$. □

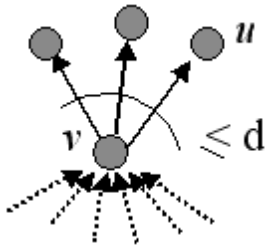

**Fig. 2.** *A vertex $v$ waits for all its neighbors $u$, which are connected to $v$ by outgoing edges $\langle v, u \rangle$.*

**Lemma 3.5.** *For $i = 1, 2, ..., p$, it holds that $\chi(G_i) \leq (\Lambda/(b \cdot p) + \Lambda/p) + 1$.*

*Proof.* Let $\mu_i$ be the following orientation of $G_i$. For each edge $e = (u, v) \in E(G_i)$, orient the edge towards the endpoint that is colored by a smaller $\varphi$-color. If $\varphi(u) = \varphi(v)$, then orient $e$ towards the endpoint with smaller Id among $u, v$. Each vertex $v$ in $G_i$ has at most $\Lambda/p$ neighbors $u$ in $G_i$ with smaller $\varphi$-colors. (This is because $\psi(u) = \psi(v)$ and $\varphi(u) < \varphi(v)$.) In addition, each vertex $v$ in $G_i$ has at most $\Lambda/(b \cdot p)$ neighbors $u$ in $G_i$ with $\varphi(v) = \varphi(u)$. Consequently, $\mu_i$ has out-degree at most $(\Lambda/(b \cdot p) + \Lambda/p)$.

Next, we prove that the orientation $\mu_i$ is acyclic. Let $C$ be a cycle of $G_i$. Let $v$ be a vertex on the cycle $C$ with the largest $\varphi$-color. If there are several vertices that satisfy this condition, let $v$ be the vertex with the greatest Id. Let $u$ and $w$ be the neighbors of $v$ in $C$. Since $\varphi(u) < \varphi(v)$ or ($\varphi(u) = \varphi(v)$ and $Id(u) < Id(v)$), the edge $(v, u)$ is oriented by $\mu_i$ towards $u$. Similarly, the edge $(v, w)$ is oriented towards $w$. Therefore, $C$ is not an oriented cycle. Consequently, $\mu_i$ is acyclic. Since the out-degree of $\mu_i$ is at most $(\Lambda/(b \cdot p) + \Lambda/p)$, Lemma 3.4 implies that $\chi(G_i) \leq (\Lambda/(b \cdot p) + \Lambda/p) + 1$. □

The next lemma shows that the family of graphs with bounded neighborhood independence is closed under taking vertex-induced subgraphs.

**Lemma 3.6.** *For a positive integer $c$, let $G = (V, E)$ be a graph with neighborhood independence at most $c$. The subgraph induced by a subset $U \subseteq V$ also has neighborhood independence at most by $c$.*

*Proof.* Let $\mathcal{G} = G(U)$ be the subgraph induced by $U$. For a vertex $u \in U$, $\Gamma_\mathcal{G}(u)$ is the neighborhood of $u$ in $\mathcal{G}$, and $\Gamma_G(u)$ is the neighborhood of $u$ in $G$. Suppose for contradiction that there exists a vertex $u \in U$ such that there is an independent set $W \subseteq \Gamma_\mathcal{G}(u)$ with cardinality $|W| > c$. For a pair of vertices $v, w \in W$, it holds that $(v, w) \notin E$. In addition, $\Gamma_\mathcal{G}(u) \subseteq \Gamma_G(u)$. Therefore, $W \subseteq \Gamma_G(u)$ is an independent set with more than $c$ vertices, and it is contained in the neighborhood $\Gamma_G(u)$ of the vertex $u$. This is a contradiction. □

We employ Lemmas 3.4 - 3.6 to proove the correctness of Procedure Defective-Color.

**Theorem 3.7.** *Suppose that Procedure Defective-Color is invoked on a graph $G$ with maximum degree $\Delta$ and with neighborhood independence bounded by a positive constant $c$. Suppose also that it receives as input three integer parameters $b \geq 1, p \geq 1, \Lambda \geq 1$, such that $b \cdot p \leq \Lambda$, $\Lambda \geq \Delta$. Then Procedure Defective-Color computes a $((\Lambda/(b \cdot p) + \Lambda/p) \cdot c + c)$-defective $p$-coloring.*

*Proof.* Recall that $G_i$ is the graph induced by vertices with $\psi$-color $i$ returned by Procedure Defective-Color, for $i = 1, 2, ..., p$. By Lemma 3.6, since $G_i$ is a subgraph of $G$, the neighborhood independence of $G_i$ is bounded by $c$. We prove that the maximum degree of $G_i$, for $i = 1, 2, ..., p$, is at most $(\Lambda/(b \cdot p) + \Lambda/p) \cdot c + c$. Suppose for contradiction that there is a vertex $v \in G_i$ such that $deg_{G_i}(v) > (\Lambda/(b \cdot p) + \Lambda/p) \cdot c + c$. Let $\varphi'$ be a legal coloring of $G_i$ that employs the minimum number of colors. Each color class of $\varphi'$ is an independent set. Therefore, for a positive integer $q$, the number of neighbors $u$ of $v$ such that $\varphi'(u) = q$ is at most $c$. Consequently, the number of different colors employed for coloring the set $\Gamma_{G_i}(v)$ of neighbors of $v$ in $G_i$ is at least $\left\lceil deg_{G_i}(v)/c \right\rceil > (\Lambda/(b \cdot p) + \Lambda/p) + 1$. However, by Lemma 3.5, $\chi(G_i) \leq (\Lambda/(b \cdot p) + \Lambda/p) + 1$, contradiction. □

We summarize this section with the following corollary.

**Corollary 3.8.** *For an integer parameter $p$, $1 \leq p \leq \Lambda$, $\Lambda \geq \Delta$, and a constant $\epsilon > 0$, a $((c + \epsilon) \cdot \frac{\Lambda}{p} + c)$-defective $p$-coloring of a graph $G$ with $I(G) \leq c$ can be computed in $O(p^2 + \log^* n)$ time.*

*Proof.* Set $b < \frac{1}{\epsilon}$. Now the corollary follows from Corollary 3.3 and Theorem 3.7. □

Observe that for graphs with bounded neighborhood independence, the product of the defect $O(\Delta/p)$ and the number of colors $p$ in the coloring produced by Corollary 3.8 is $O(\Delta)$. This is in sharp contrast to the current state-of-the-art for distributed defective coloring in *general* graphs [4, 19], which is $O(\Delta/p)$-defective $p^2$-coloring. On the other hand, the latter coloring can be computed faster, specifically, within $O(\log^* n)$ time [19].

## 4 Legal Coloring Graphs with Bounded Neighborhood Independence

In this section we employ the defective coloring algorithm from the previous section for legal vertex coloring of graphs with neighborhood independence at most $c$. We start with presenting a simpler algorithm that does not achieve our strongest bounds (Section 4.1). Then we proceed to improving these bounds further (Section 4.2).

### 4.1 The main algorithm

In this section we employ the defective coloring algorithm from the previous section for legal vertex coloring of graphs with neighborhood independence at most $c = O(1)$. Fix an arbitrarilly small constant $\epsilon > 0$. Once a $(c + \epsilon) \cdot \frac{\Delta}{p} + c = O(\Delta/p)$-defective $p$-coloring $\psi$ of a graph $G$ is computed, it constitutes a vertex partition $V_1, V_2, ..., V_p$, in which $V_i$ is the set of vertices with $\psi$-color $i$, for $i = 1, 2, ..., p$. In other words, $V = \bigcup_{i=1}^n V_i$, and for a pair of distinct indices $i, j \in \{1, 2, ..., p\}$, $i \neq j$, $V_i \cap V_j = \emptyset$. The subgraph $G_i$ induced by $V_i$ has maximum degree $O(\Delta/p)$, since each vertex in $G$ has at most $O(\Delta/p)$ neighbors with the same $\psi$-color. Therefore, one can legally color the subgraphs $G_1, G_2, ..., G_p$ employing $O(\Delta/p)$ colors for each subgraph, using Lemma 2.1(2). These colorings, denoted $\varphi_1, \varphi_2, ..., \varphi_p$, are computed in parallel on the subgraphs $G_1, G_2, ..., G_p$. Let $m = O(\Delta/p)$ denote the maximum number of colors employed by $\varphi_i$, for $i = 1, 2, .., p$. Next, the colorings are combined into a unified legal coloring $\varphi$ of $G$ as follows. Observe that each vertex $v \in V$ belongs exactly to one subgraph $G_j$ among $G_1, G_2, ..., G_p$. We set $\varphi(v) = \varphi_j(v) + (j-1) \cdot m$. (The color $\varphi(v)$ can also be thought of as a pair $(j, \varphi_j(v))$, where $v \in V_j$.)

The coloring $\varphi$ is a legal coloring of $G$, since for any pair of vertices $u, v \in G$, if they belong to the same subgraph $G_i$, $i \in \{1, 2, ..., p\}$, then $\varphi_i(v) \neq \varphi_i(u)$, and, therefore, $\varphi(v) \neq \varphi(u)$. Otherwise, $u$ belongs to $G_i$, and $v$ belong to $G_j$, for some $1 \leq i \neq j \leq p$. Since $|(j-1) \cdot m - (i-1) \cdot m| \geq m$, and $|\varphi_j(u) - \varphi_i(v)| \leq m - 1$, in this case also it holds that $\varphi(v) \neq \varphi(u)$.

The running time required for computing $\varphi$ is the running time of computing a legal coloring of a graph with degree $\left\lfloor (c+\epsilon) \cdot \frac{\Delta}{p} + c \right\rfloor = O(\Delta/p)$. Hence, by Lemma 2.1(2), the running time of computing $\varphi$ from a given $O(\Delta/p)$-defective $p$-coloring $\psi$ is $O(\Delta/p + \log^* n)$. By Corollary 3.8, the running time of computing $\psi$ is $O(p^2 + \log^* n)$. Therefore, the overall running time of computing $\varphi$ from scratch is $O(\Delta/p + p^2 + \log^* n)$. This running time is optimized by setting $p = \lfloor \Delta^{1/3} \rfloor$, resulting in overall $O(\Delta^{2/3} + \log^* n)$ time. The number of colors employed by the resulting legal coloring is at most $((c+\epsilon) \cdot \frac{\Delta}{p} + c + 1) \cdot p \leq (2c + \epsilon') \cdot \Delta$, for any constant $\epsilon'$, $\epsilon' > \epsilon$. We summarize this result in the following lemma.

**Lemma 4.1.** *For any constant $\epsilon' > 0$, a legal $((2c + \epsilon') \cdot \Delta)$-coloring of graphs with neighborhood independence at most $c$ can be computed in $O(\Delta^{2/3} + \log^* n)$ time.*

Next, we present a significantly faster $O(\Delta)$-coloring procedure for the family of graphs with neighborhood independence bounded by $c$, for a positive constant $c$. The procedure is called *Procedure Legal-Color*. During its execution defective colorings are computed several times. In the first phase of the procedure a defective coloring of the input graph is computed. This coloring forms a partition of the original graph into vertex-disjoint subgraphs, each with maximum degree smaller than $\Delta$. Then the procedure is invoked recursively on these subgraphs in parallel. This invocation partitions each subgraph into more subgraphs with yet smaller maximum degrees. This process repeats itsel until the maximum degrees of all subgraphs are sufficiently small. Then, legal colorings of these subgraphs are computed in parallel, and merged into a unified legal coloring of the input graph. Even though Procedure Defective-Color is invoked many times by Procedure Legal-Color, the running time of Procedure Legal-Color is much smaller than the time given in Lemma 4.1. The improvement in time is achieved by selecting different parameters in the defective colorings computations, making the invocations significantly faster than a single invocation with the parameter $p = \lfloor \Delta^{1/3} \rfloor$.

Procedure Legal-Color receives as input a graph $G$, positive integer parameters $b, p$ such that $p > 4c$, $1 \leq b \cdot p \leq \Delta$, a parameter $\lambda$ such that $2c < \lambda \leq \Delta$, and a parameter $\Lambda$. The parameter $\Lambda$ represents an upper bound on the maximum degree of the input graph. Initially, $\Lambda$ is set to $\Delta$. Later as the procedure is invoked recursively, this parameter is set to smaller values, and it keeps decreasing as the recursion proceeds. The threshold parameter $\lambda$ determines the termination condition of the recursion. Specifically, if $\Lambda \leq \lambda$ then a legal $(\Lambda + 1)$-coloring of $G$ is computed directly using Lemma 2.1(2). Otherwise, an $O(\Delta/p)$-defective $p$-coloring $\psi$ of $G$ is computed, producing the subgraphs $G_1, G_2, ..., G_p$ induced by the $\psi$-color classes $1, 2, ..., p$, respectively. Let $\lambda' = O(\Delta/p)$ denote the defect of the coloring $\psi$. Next, Procedure Legal-Color is invoked on each of these subgraphs recursively, with the degree parameter $\lambda'$. All other parameters, that is, $b, p$, and $\lambda$, do not change throughout the recursion.

For technical convenience, Procedure Legal-Color returns not only the resulting coloring $\varphi$, but also an upper bound $\vartheta$ on the number of colors that this coloring employs. On the bottom level of the recursion, i.e., when $\Lambda \leq \lambda$, the number of colors that is used by $\varphi$ is at most $\Lambda + 1$. In this case the algorithm (see line 3) sets $\vartheta = \Lambda + 1$. In the more general case when $\Lambda$ is greater than the threshold value $\lambda$, the algorithm invokes itself recursively on each of the subgraphs $G_1, G_2, ..., G_p$. These recursive invocations return the pairs $(\varphi_1, \vartheta_1), (\varphi_2, \vartheta_2), ..., (\varphi_p, \vartheta_p)$, where for each $i = 1, 2, ..., p$, $\varphi_i$ is a $\vartheta_i$-coloring of $G_i$. These colorings are merged into a unified $\vartheta$-coloring $\varphi$ of the entire graph $G$, with $\vartheta = \Sigma_{i=1}^{p} \vartheta_i$. It will be shown later that, in fact, $\vartheta_i = \vartheta_j$ for every $i, j \in \{1, 2, ..., p\}$. (In other words, the *upper bound* on the number of colors employed by the coloring $\varphi_i$ that the algorithm returns is equal to the *upper bound* that it returns for the coloring $\varphi_j$. This does not necessarily mean that the two coloring use exactly the same number of colors.) This implies that one can actually set $\vartheta = \vartheta' \cdot p$, where $\vartheta' = \vartheta_i$ for some $i \in \{1, 2, ..., p\}$, as the algorithm indeed does in line 11.

Moreover, to obtain the unified legal coloring $\varphi$ the algorithm just adds to the color $\varphi_i(v)$ of a vertex $v \in V_i = V(G_i)$ the value $(i-1) \cdot \vartheta_i$. Since $\vartheta_i = \vartheta'$ for every $i \in \{1, 2, ..., p\}$, it follows that this way vertices of $V_i$ end up being $\varphi$-colored by a color from the palette $\{(i-1)\vartheta' + 1, (i-1)\vartheta' + 2, ..., i \cdot \vartheta'\}$. Consequently, for two vertices $u, w$, $u \in V_i$, $w \in V_j$, $i \neq j$, their $\varphi$-colors belong to disjoint palettes, and are, thus, different. Below we provide the pseudocode of Procedure Legal-Color.

---

**Algorithm 2** Procedure Legal-Color($G, b, p, \lambda, \Lambda$ )

An algorithm for each vertex $v \in V$.

1: **if** ($\Lambda \leq \lambda$) **then**
2:     $\varphi :=$ a legal $(\Lambda + 1)$-coloring of $G$ using Lemma 2.1(2)
3:     $\vartheta := \Lambda + 1$    /* number of colors employed by $\varphi$ */
4: **else**
5:     $\psi :=$ Defective-Color($G, b, p, \Lambda$)
    /* for $i = 1, 2, ..., p$, let $G_i$ be the graph induced by vertices of $\psi$-color $i$ */
6:     $\Lambda' := \lfloor (\Lambda/(b \cdot p) + \Lambda/p) \cdot c + c \rfloor$    /* defect parameter of $\psi$ */
7:     **for** $i = 1, 2, ..., p$, in parallel **do**
8:        $(\varphi_i, \vartheta') :=$ Legal-Color($G_i, b, p, \lambda, \Lambda'$)
       /* recursive invocation that computes $\vartheta'$-coloring $\varphi_i$ of $G_i$ */
9:        **if** $v \in V(G_i)$ **then**
10:           $\varphi(v) := \varphi_i(v) + (i-1) \cdot \vartheta'$
11:           $\vartheta := \vartheta' \cdot p$
12:        **end if**
13:     **end for**
14: **end if**
15: **return** $(\varphi, \vartheta)$    /* return the legal coloring and the number of employed colors */

---

Let $\epsilon$ be an arbitrarily small positive constant. We execute Procedure Legal-Color on the input graph $G$ of maximum degree $\Delta$ and neighborhood independence bounded by $c$, with the parameters $b = \lceil \Delta^{\epsilon/6} \rceil, p = \lceil \Delta^{\epsilon/3} \rceil, \lambda = \lceil \Delta^{\epsilon} \rceil, \Lambda = \Delta$. In the next lemma we analyze the running time of this invocation.

**Lemma 4.2.** *Procedure Legal-Color invoked with the parameters* $b = \lceil \Delta^{\epsilon/6} \rceil, p = \lceil \Delta^{\epsilon/3} \rceil, \lambda = \lceil \Delta^{\epsilon} \rceil, \Lambda = \Delta$ *terminates in* $O(\Delta^{\epsilon} + \log^* n)$ *time.*

*Proof.* In each recursion level it holds that

$$\Lambda' = \lfloor (\Lambda/(b \cdot p) + \Lambda/p) \cdot c + c \rfloor \leq (\Lambda/\Delta^{\epsilon/2} + \Lambda/\Delta^{\epsilon/3}) \cdot c + c \leq 3 \cdot c \cdot \Lambda/\Delta^{\epsilon/3}. \tag{1}$$

Therefore, for a sufficiently large $\Delta$, the number of recursion levels $r$ satisfies

$$r \leq \log_{\Delta^{\epsilon/3}/3c} \Delta = \frac{\log \Delta}{\log(\Delta^{\epsilon/3}/3c)} = \frac{\log \Delta}{(\epsilon/3) \cdot \log \Delta - \log 3c} = O(1). \tag{2}$$

(Recall that both $\epsilon$ and $c$ are constants.)

By Lemma 3.3, each recursion level, except for the last one, requires $O(\Delta^{\epsilon} + \log^* n)$ time. (This running time is dominated by the time required for executing Procedure Defective-Color in line 5 of Algorithm 2.) By Lemma 2.1(2), the last recursion level, in which $\Lambda \leq \lambda$, requires $O(\lambda + \log^* n) = O(\Delta^{\epsilon} + \log^* n)$ time. Therefore, the overall running time is also $O(\Delta^{\epsilon} + \log^* n)$. □

We remark that equation (1) in the proof of Lemma 4.2 holds for any (not necessarily constant) $\epsilon$, $0 < \epsilon < 1$. On the other hand, for constant $\epsilon > 0$ this equation can be refined.

**Lemma 4.3.** *For any constant $\eta > 0$, for a sufficiently large $\Delta$ it holds that $\Lambda' \le c \cdot (1 + \eta) \cdot \Lambda / \Delta^{\epsilon/3}$.*

*Proof.* Observe that $(\Lambda/\Delta^{\epsilon/2} + \Lambda/\Delta^{\epsilon/3}) \cdot c + c = (\Lambda/\Delta^{\epsilon/3}) \cdot (1 + 1/\Delta^{\epsilon/6}) \cdot c + c$.
Recall also that $\Lambda \ge \Delta^{\epsilon}$, and thus $\Lambda/\Delta^{\epsilon/3} \ge \Delta^{2\epsilon/3}$. For $\Delta > (2/\eta)^{6/\epsilon}$ it holds that $1/\Delta^{\epsilon/6} < \eta/2$, and so

$$(\Lambda/\Delta^{\epsilon/3}) \cdot (1 + 1/\Delta^{\epsilon/6}) \cdot c + c \le (\Lambda/\Delta^{\epsilon/3}) \cdot (1 + \eta/2) \cdot c + c. \tag{3}$$

Also, $c < c \cdot (\eta/2) \cdot \Delta^{\epsilon/6} < c \cdot (\eta/2) \Delta^{2\epsilon/3} \le c \cdot (\eta/2) \cdot (\Lambda/\Delta^{\epsilon/3})$, and so, the right-hand-side of equation (3) is at most $c \cdot (1 + \eta) \cdot \Lambda/\Delta^{\epsilon/3}$, completing the proof. □

In the next lemma we analyze the number of colors used by the coloring $\varphi$, which is returned by the procedure.

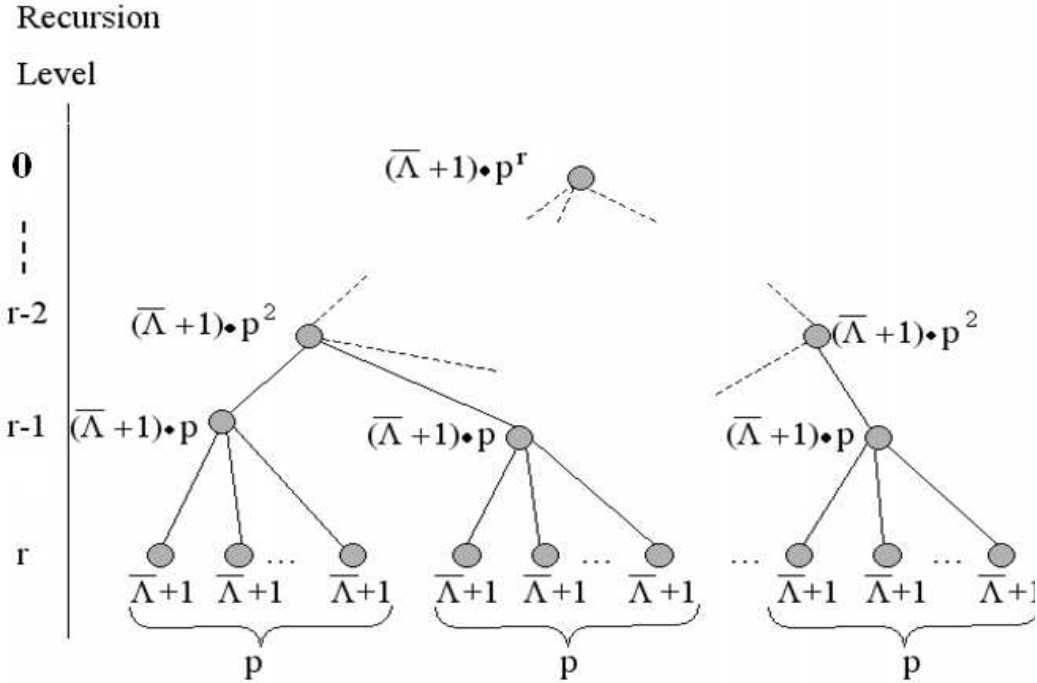

**Fig. 3.** *The recursion tree. For each node $\alpha$, the value that appears near $\alpha$ in the figure represents the number of colors that are employed by the recursive invocation that corresponds to $\alpha$.*

**Lemma 4.4.** *Procedure Legal-Color invoked with the parameters $b = \lceil \Delta^{\epsilon/6} \rceil, p = \lceil \Delta^{\epsilon/3} \rceil, \lambda = \lceil \Delta^{\epsilon} \rceil, \Lambda = \Delta$ computes a legal $O(\Delta)$-coloring.*

*Proof.* Consider the recursion tree $\tau$ of Procedure Legal-Color invoked with the parameters $(G, b, p, \lambda, \Lambda)$, set as above. (See Figure 3.) With each node $\alpha$ of $\tau$ we associate the value $\Lambda_\alpha$ of the parameter $\Lambda$ in the recursive invocation corresponding to $\alpha$. By induction on the level $i$ in the tree $\tau$ it can be shown that all nodes $\alpha$ of $\tau$ that have level $i$ have the same value $\Lambda_\alpha = \Lambda^{(i)}$. The choice of parameters also guarantees that $\Lambda = \Lambda^{(0)} > \Lambda^{(1)} > \Lambda^{(2)} > \dots$. Let $r$ be the smallest index such that $\Lambda^{(r)} \le \lambda$. Denote $\Lambda^{(r)} = \hat{\Lambda}$. Observe that nodes $\alpha$ of level $r$ in $\tau$ are the leaves of $\tau$, and every leaf of $\tau$ has level $r$. Hence all leaves of $\tau$ have the same level. Moreover, it follows that in each of theses leaf-invocations $\alpha$ of $\tau$, the corresponding subgraph $G[\alpha]$ is colored with at most $\Lambda^{(r)} + 1$ colors, and that each of these invocations returns $\vartheta^{(r)} = \Lambda^{(r)} + 1$. Hence, by induction on $j$, $j = 1, 2, ..., r$, for any two invocations $\alpha$ and $\beta$ of the $j$'th level in the recursion tree $\tau$, these two invocations return the same values of $\vartheta$. (The induction base is the case $j = r$, and in the step we proceed from $j + 1$ to $j$, for $j \in \{2, 3, ..., r\}$.)

For $j \in \{1, 2, ..., r\}$, denote by $\vartheta^{(j)}$ the parameter $\vartheta$ returned by some invocation $\alpha$ of level $j$. Next, we argue, again by the induction on $i$ with the base case $j = r$, that the $\vartheta$-coloring $\varphi$ returned by the invocation $\alpha$ is legal. The base case follows from Lemma 2.1(2). For the induction step let $G[\alpha]$ denote the subgraph of $G$ on which the invocation of Procedure Legal-Color that corresponds to the node $\alpha$ is invoked. Also, denote by $G_1[\alpha], G_2[\alpha], ..., G_p[\alpha]$ the $p$ subgraphs of $G[\alpha]$ that correspond to the $p$ children of $\alpha$. We remark that, in general, line 5 of Procedure Legal-Color may partition the graph $G[\alpha]$ into less than $p$ subgraphs. In this case, however, the routine leaves $\vartheta^{(j+1)}$ empty color classes for each of the "missing" subgraphs. It does so by setting $\vartheta^{(j)} = p \cdot \vartheta^{(j+1)}$, even though it really needs a smaller palette. These redundant colors help to maintain uniform bounds on the number of colors used to color subgraphs of the same recursion level.

For each $i \in \{1, 2, ..., p\}$, vertices of $G_i[\alpha]$ are colored by colors from the palette $\{(i-1) \cdot \vartheta^{(j+1)} + 1, (i-1) \cdot \vartheta^{(j+1)} + 2, i \cdot \vartheta^{(j+1)}\}$. Thus, for a pair of vertices $u, w$, $u \in G_i[\alpha]$, $w \in G_{i'}[\alpha]$, $i \neq i'$, $i, i' \in \{1, 2, ..., p\}$, the invocation $\alpha$ colors them by distinct colors. (Denote these colors by $\varphi[\alpha](u)$ and $\varphi[\alpha](w)$, respectively. Then $\varphi[\alpha](u) \neq \varphi[\alpha](w)$.) Consider also the case that $u, w \in G_i[\alpha]$ belong to the same subgraph $G_i[\alpha]$ of $G_i$, and $(u, w) \in E$. By the induction hypothesis, the coloring $\varphi_i[\alpha]$ of $G_i[\alpha]$ returned by the $i$'th child invocation of $\alpha$ is legal. Hence $\varphi_i[\alpha](u) \neq \varphi_i[\alpha](w)$. Recall that $\varphi[\alpha](u) = \varphi_i[\alpha](u) + (i-1) \cdot \vartheta^{(i+1)}$ and $\varphi[\alpha](w) = \varphi_i[\alpha](w) + (i-1) \cdot \vartheta^{(i+1)}$, and so $\varphi[\alpha](u) \neq \varphi[\alpha](w)$, as required.

Finally, we provide an estimate on the number of colors $\vartheta = \vartheta^{(0)}$ employed by the coloring $\varphi$ returned by the root invocation Legal-Color$(G, b, p, \lambda, \Lambda)$ of the recursion tree $\tau$. We have already shown that $\vartheta^{(r)} = \Lambda^{(r)} = \hat{\Lambda} + 1$. Also, for any level $j$, $0 \leq j \leq r - 1$, $\vartheta^{(j)} = p \cdot \vartheta^{(j+1)}$. Hence the overall number of colors satisfies $\vartheta^{(0)} \leq \vartheta^{(r)} \cdot p^r = (\hat{\Lambda} + 1) \cdot p^r$. By equation (2) in proof of Lemma 4.2, it holds that

$$r \leq \frac{\log(\Delta/\hat{\Lambda})}{\log(\Delta^{\epsilon/3}/3c)} = O(1).$$

(This is true for a constant $\epsilon > 0$.) Finally, for a sufficiently large $\Delta$,

$$\begin{aligned}
(\hat{\Lambda}+1) \cdot p^r &\leq& (\hat{\Lambda}+1) \cdot \left[\Delta^{\epsilon/3}\right]^r &\leq& (\hat{\Lambda}+1) \cdot (3c+1)^r \cdot (\frac{\Delta^{\epsilon/3}}{3c})^r \\
&\leq& (\hat{\Lambda}+1) \cdot (3c+1)^r \cdot (\frac{\Delta^{\epsilon/3}}{3c})^{\frac{\log(\Delta/\hat{\Lambda})}{\log(\Delta^{\epsilon/3}/3c)}} &=& (3c+1)^r \cdot (\hat{\Lambda}+1) \cdot (\Delta/\hat{\Lambda}) &=& O(\Delta),
\end{aligned}$$

since $c, r = O(1)$, and $\frac{\hat{\Lambda}+1}{\hat{\Lambda}} \leq 2$. Hence $\vartheta^{(0)} = O(\Delta)$, completing the proof. $\square$

We summarize the properties of Procedure Legal-Color in the following Theorem.

**Theorem 4.5.** *For an arbitrarily small positive constant $\epsilon$, our algorithm computes an $O(\Delta)$-coloring of graphs with bounded neighborhood independence, in time $O(\Delta^\epsilon + \log^* n)$.*

Next we show that one can compute a legal coloring of $G$ much faster, at the expense of increasing the number of employed colors. To this end, one needs only to select different parameters for the invocation of Procedure Legal-Color.

**Theorem 4.6.** *For an arbitrarily small positive constant $\eta$, our algorithm computes an $O(\Delta^{1+\eta})$-coloring of graphs with bounded neighborhood independence, in time $O(\log \Delta \log^* n)$.*

*Proof.* Let $t > 2$ be an arbitrarily large constant. Set $\lambda = (3c+1)^{6t}, b = \lambda^{1/3} = (3c+1)^{2t}, p = \lambda^{1/6} = (3c+1)^t, \Lambda = \Delta$. In each recursion level it holds that

$$\Lambda' = \lfloor (\Lambda/(b \cdot p) + \Lambda/p) \cdot c + c \rfloor \leq (\Lambda/(3c+1)^{3t} + \Lambda/(3c+1)^t) \cdot c + c \leq 3 \cdot c \cdot \Lambda/(3c+1)^t.$$

Therefore, the number of recursion levels in this case is at most $r = \log_{(3c+1)^{t-1}} \Delta$. By Corollary 3.3, invoking Procedure Defective-Color on each level requires $O((3c+1)^{6t} + \log^* n) = O(\log^* n)$ time. Other steps of Algorithm 2, except for the recursive invocation step, are executed locally, and require zero time. Hence each recursion level (except for the bottom level) requires $O(\log^* n)$ time. On the bottom level of the recursion we invoke a $(\Lambda + 1)$-coloring algorithm from [4] on subgraphs of maximum degree at most $\Lambda$. (See Lemma 2.1(2).) This algorithm requires $O(\Lambda + \log^* n)$ time. Note that at the bottom line of the recursion $\Lambda \leq \lambda$, and $\lambda = (3c+1)^t = O(1)$ is a constant. Hence this running time is $O(\log^* n)$ as well. We conclude that the overall running time is $O(\log_{(3c+1)^{t-1}} \Delta \cdot \log^* n) = O(\log \Delta \log^* n)$.

The inductive proof of Lemma 4.4 is applicable as is also for the new selection of parameters. Hence, the produced coloring is legal. The number of colors in this case is at most

$$\begin{aligned}(\lambda + 1) \cdot p^r &= ((3c+1)^{6t} + 1) \cdot (3c+1)^{(t-1)\cdot r} \cdot (3c+1)^r \leq ((3c+1)^{6t} + 1) \cdot \Delta \cdot \Delta^{1/(t-1)} \\ &= O(\Delta^{1+1/(t-1)}).\end{aligned}$$

Given a constant $\eta > 0$, set $t$ to be sufficiently large so that $1/(t-1) < \eta$. Hence the number of colors is at most $O(\Delta^{1+\eta})$. □

### 4.2 An improved version

In this section we show that our algorithms can be modified to guarantee running time of $O(\log \Delta \cdot \frac{\log^* \Delta}{\log(\log^* \Delta)}) + \frac{1}{2} \log^* n$, while maintaining the same bound of $\Delta^{1+\eta}$ on the number of colors. We do so in two steps. First, we prove a bound of $O(\log \Delta \log^* \Delta + \log^* n)$, and then we improve it further.

The change that we introduce to our algorithms is the following one. Before invoking Procedure Legal-Color we invoke Linial's algorithm (see Lemma 2.1(1)) for computing a legal $O(\Delta^2)$-coloring $\rho$ of the original graph $G$. This invocation requires $O(\log^* n)$ time. This auxiliary coloring will help us to execute Procedure Legal-Color faster. Specifically, we will use the following result of [19].

**Theorem 4.7.** *[Theorem 4.9 in [19]]: Given a $d'$-defective $M$-coloring of a graph $G$ of maximum degree $\Delta$ and a defect parameter $d$, $d' \leq d < \Delta$, a $d$-defective $O((\frac{\Delta - d'}{d+1-d'})^2)$-coloring of $G$ can be computed in $O(\log^* M)$ rounds.*

We will use this theorem with $d' = 0$. In this case the resulting coloring is a $d$-defective $O((\frac{\Delta}{d})^2)$-coloring. Given a parameter $p$, set $d = \lfloor \Delta/p \rfloor$. We get a $\lfloor \Delta/p \rfloor$-defective $O(p^2)$-coloring, in $O(\log^*(\Delta^2)) = O(\log^* \Delta)$ time. Our algorithm invokes Procedure Defective-Color (on line 5 of Algorithm 1) in each recursive invocation of Procedure Legal-Color. Procedure Defective-Color, in turn, computes a $\lfloor \frac{\Delta}{b \cdot p} \rfloor$-defective $O((b \cdot p)^2)$-coloring $\varphi$ on line 1 of Algorithm 1, using Lemma 2.1(3). This step requires $O(\log^* n)$ time. In the modified version of our algorithm we will use the algorithm of Theorem 4.7 with tha auxiliary $O(\Delta^2)$-coloring $\rho$ as its input, and with the parameter $d = \lfloor \frac{\Delta}{b \cdot p} \rfloor$. In this way we obtain the desired $\lfloor \frac{\Delta}{b \cdot p} \rfloor$-defective $O((b \cdot p)^2)$-coloring $\varphi$ in $O(\log^* \Delta)$ additional time, instead of $O(\log^* n)$ time.

Since the number of recursion levels is $r = O(\log \Delta)$, it follows that the running time of all recursion levels of Procedure Legal-Color except for the bottom level is $O(\log \Delta \log^* \Delta)$. On the bottom level of the recursion we invoke an algorithm from Lemma 2.1(2) (due to [4]) for computing a legal $(\Lambda + 1)$-coloring of a graph with maximum degree at most $\Lambda$. This algorithm requires $O(\Lambda + \log^* n)$ time. However, at this point $\Lambda \leq \lambda = O(1)$, and thus this step requires $O(\log^* n)$ time. Hence the total running time of all recursion levels is $O(\log \Delta \log^* \Delta + \log^* n)$. Computing of the auxiliary coloring $\rho$ requires additional $O(\log^* n)$ time. Hence the overall running time of the algorithm is $O(\log \Delta \log^* \Delta + \log^* n)$.

This estimate can be slightly improved by setting parameters of Procedure Legal-Color differently. Specifically, set $\lambda = \log^* \Delta, p = \lambda^{1/6}, b = \lambda^{1/3}, \Lambda = \Delta$. The depth $r$ of the recursion tree $\tau$ becomes $r = O(\frac{\log \Delta}{\log \lambda}) = O(\frac{\log \Delta}{\log(\log^* \Delta)})$. The running time of each recursion level except for the bottom level is still dominated by the running time of Procedure Defective-Color, i.e, $O((b \cdot p)^2 + \log^* \Delta) = O(\log^* \Delta)$. On the bottom level of the recursion computations of $(\lambda + 1)$-colorings of graphs with maximum degrees at most $\lambda$ require $O(\lambda + \log^* n)$ time. Since $\lambda = \log^* \Delta$, this is $O(\log^* n)$ as well. Hence the overall running time of the algorithm becomes $O(\log \Delta \cdot \frac{\log^* \Delta}{\log(\log^* \Delta)} + \log^* n)$.

Another notable point on the tradeoff curve is $\Delta^{1+o(1)}$-coloring in $O((\log \Delta)^{1+\eta} + \log^* n)$ time, for a fixed arbitrarily small constant $\eta > 0$. To achieve this we set $\lambda = \log^\eta \Delta, b = \lambda^{1/3}, p = \lambda^{1/6}$. The depth of the recursion tree becomes $r = \frac{\log(\Delta/\log^\eta \Delta)}{\log((\log^{\eta/6} \Delta)/3c)} = O(\frac{\log \Delta}{\log \log \Delta})$. (Recall that $c$ is a constant bound on the graph neighborhood independence.) The running time of each recursion level except for the last one is $O((bp)^2 + \log^* \Delta) = O(\log^\eta \Delta)$ (by Corollary 3.3). The running time of the last recursion level is $O(\lambda + \log^* \Delta) = O(\log^\eta \Delta)$ as well. Hence the overall running time is $O(r \cdot \log^\eta \Delta) + O(\log^* n) = O(\log^{1+\eta} \Delta + \log^* n)$. The overall number of colors is $O(\lambda \cdot p^r) = O(\Delta) \cdot (3c)^r = \Delta^{1+o(1)}$.

Note also that the analysis above implies that the leading constant factor in front of $\log^* n$ is just 2. Specifically, one $\log^* n$ term is required for computing the auxiliary coloring $\rho$, and another $\log^* n$ term is used for computing $(\lambda + 1)$-coloring of subgraphs with maximum degree at most $\lambda$ at the bottom level of the recursion. However, a close inspection of the algorithm of [4] reveals that it spends $\log^* n$ time for computing an $O(\lambda^2)$-coloring $\rho'$, and then spends another $O(\lambda)$ time to convert it into a $(\lambda + 1)$-coloring. On the other hand, given an auxiliary $O(\Delta^2)$-coloring $\rho$, one can convert it into an $O(\lambda^2)$-coloring $\rho'$ of a subgraph with maximum degree at most $\lambda$ in $O(\log^* \Delta)$ time. (This can be done using Linial's algorithm [21].) Hence the running time of our algorithm for computing $O(\Delta^{1+\eta})$-coloring reduces to $O(\log \Delta \frac{\log^* \Delta}{\log(\log^* \Delta)}) + \log^* n$. Moreover, one can compute the auxiliary coloring $\rho$ using an algorithm of Szegedy and Vishwanathan [27] (instead of Linial's algorithm [21]) in $\frac{1}{2} \log^* n$ time. This leads to the bound of $O(\log \Delta \frac{\log^* \Delta}{\log(\log^* \Delta)}) + \frac{1}{2} \log^* n$ on the running time of our algorithm. These considerations enable one to replace $O(\log^* n)$ by $\frac{1}{2} \log^* n$ also in Theorem 4.5. To summarize:

**Theorem 4.8.** *For any constant $\epsilon > 0$, and a graph $G$ with bounded neighborhood independence,*
*(1) an $O(\Delta)$-coloring of $G$ can be computed in $O(\Delta^\epsilon) + \frac{1}{2} \log^* n$ time.*
*(2) an $O(\Delta^{1+\epsilon})$-coloring of $G$ can be computed in $O(\log \Delta \frac{\log^* \Delta}{\log(\log^* \Delta)}) + \frac{1}{2} \log^* n$ time.*
*(3) a $\Delta^{1+o(1)}$-coloring of $G$ can be computed in $O((\log \Delta)^{1+\epsilon}) + \frac{1}{2} \log^* n$ time.*

## 5 Legal Edge Coloring in General Graphs

In this section we show that the techniques described in Sections 3 and 4 can be used to devise very efficient edge coloring algorithms for *general* graphs. We start with the observation that for any graph $G$, its line graph has neighborhood independence at most 2.

**Lemma 5.1.** *For a graph $G = (V, E)$, the line graph $L(G)$ has neighborhood independence bounded by 2.*

*Proof.* Let $u$ be a vertex in $L(G)$ with at least three neighbors. Let $e_u \in E$ be the corresponding edge of $u$. Let $v, w, x$ be any three neighbors of $u$ in $L(G)$. Let $e_v, e_w, e_x$ be the corresponding edges in $E$, respectively. Each edge $e_v, e_w, e_x$ share a common endpoint with $e_u$. Therefore, at least two edges among $e_v, e_w, e_x$ share a common endpoint. Suppose, without loss of generality, that these are $e_v$ and $e_w$. Consequently, the vertices $v$ and $w$ are neighbors in $L(G)$. Hence the neighborhood $\Gamma(u)$ of $u$ does not contain three independent vertices. Consequently, the neighborhood independence of $L(G)$ is at most 2. □

Observe that Lemma 5.1 extends directly to line graphs of general $r$-hypergraphs. Specifically, for any hypergraph $\mathcal{H}$, the neighborhood independence of the line graph $L(\mathcal{H})$ is at most $r$. It follows that our results for graphs of bounded neighborhood independence (Theorem 4.8) apply to line graphs of $r$-hypergraphs, for any constant positive integer $r$.

Observe that, by definition, for any graph $G$ and positive integer $k$, a legal $k$-coloring of *vertices* of $L(G)$ is a legal $k$-coloring of *edges* of $G$, and vice versa. Note also that for an edge $e = (u, w)$ in $G$, the number of edges incident to it is $(deg(u) - 1) + (deg(w) - 1)$. Hence the maximum degree $\Delta(L(G))$ of the line graph $L(G)$ satisfies $\Delta(L(G)) \leq 2(\Delta - 1)$, where $\Delta = \Delta(G)$. Consequently, if we are given a line graph $L(G)$ of a graph $G$ with $\Delta(G) = \Delta$, our algorithm can compute an $O(\Delta(L(G))) = O(\Delta)$-vertex-coloring of $L(G)$ in $O(\Delta^\epsilon) + \frac{1}{2}\log^* n$ time, for any constant $\epsilon > 0$. Similarly, one can also compute $O(\Delta^{1+\eta})$-vertex-coloring (respectively, $\Delta^{1+o(1)}$-vertex-coloring) of $L(G)$ in $O(\log \Delta \frac{\log^* \Delta}{\log(\log^* \Delta)}) + \frac{1}{2}\log^* n$ (resp., $(\log \Delta)^{1+\zeta} + \frac{1}{2}\log^* n$) time, for $\eta, \zeta > 0$ being arbitrarily small positive constants. These vertex colorings give rise directly to edge coloring of $G$ with the same number of colors.

On the other hand, in the distributed edge-coloring problem we are given as input the graph $G$, rather than its line graph $L(G)$. To overcome this difficulty we can simulate the distributed computation of an algorithm on $L(G)$ using the network $G = (V, E)$.

**Lemma 5.2.** *Any algorithm with running time $T$ for the line graph $L(G)$ of the input graph $G$, can be simulated by $G$, and requires at most $2T + O(1)$ time.*

*Proof.* For each edge $e$ in $E$, one of the endpoints of $e$ simulates a vertex in $L(G)$ correspond to $e$. (Hence, each vertex in $G$ may simulate many vertices of $L(G)$.) Specifically, for each edge $e = (u, v) \in E$, such that $Id(u) < Id(v)$, the vertex that corresponds to $e$ in $L(G)$ is simulated by $u$. We denote the vertex in $L(G)$ that corresponds to $e$ by $v_e$. The $Id$ of $v_e$ is set as the ordered pair $\langle Id(u), Id(v) \rangle$. This guarantees unique $Id$s for vertices in $L(G)$. Sending a message from a vertex $w$ in $L(G)$ to its neighbor $w'$ is simulated as follows. If the vertices that simulate $w$ and $w'$ are neighbors in $G$, the message is sent directly. Otherwise, the distance between the simulating vertices is 2. The vertices $w$ and $w'$ correspond to edges $e$ and $e'$ in $E$ that share a common endpoint $v'$. In this case, the message is sent from the vertex that simulates $w$ to $v'$, and from $v'$ to the vertex that simulates $w$. Consequently, any algorithm for the line graph can be simulated on the original graph, increasing the running time by a factor of at most 2. The additive term of $O(1)$ in the running time above reflects the time spent for computing unique edge identifiers. □

Now we apply Lemma 5.2 in conjunction with our results for vertex-coloring of $L(G)$, and obtain the following theorem.

**Theorem 5.3.** *For a graph $G = (V, E)$ with maximum degree $\Delta$, and positive arbitrarily small constants $\epsilon, \eta, \zeta > 0$, our algorithm computes: (1) $O(\Delta)$-edge-coloring of $G$ in $O(\Delta^\epsilon) + \frac{1}{2}\log^* n$ time.*
*(2) $O(\Delta^{1+\eta})$-edge-coloring of $G$ in $O(\log \Delta \frac{\log^* \Delta}{\log(\log^* \Delta)}) + \frac{1}{2}\log^* n$ time.*
*(3) $\Delta^{1+o(1)}$-edge-coloring of $G$ in $O((\log \Delta)^{1+\zeta}) + \frac{1}{2}\log^* n$ time.*

Despite its simplicity and generality, the simulation technique that we described above has some downsides. Most notably, it causes tha algorithm to send up to $\Delta$ messages through a single edge in a single round. Consequently, the resulting algorithm requires message size of $O(\Delta \log n)$. In what follows we present an edge-coloring variant of our algorithm from Sections 3 and 4. This algorithm provides nearly the same (and actually, in case (2) slightly better) bounds as in Theorem 5.3, but does so using much shorter messages. Specifically, in case (2) we eliminate the slack of $\frac{\log^* \Delta}{\log(\log^* \Delta)}$, and obtain the running time of $O(\log \Delta) + \log^* n$ for $O(\Delta^{1+\eta})$-coloring. On the other hand, the $\frac{1}{2}\log^* n$ term in all the three cases will become $\log^* n$.

One ingredient of our edge-coloring algorithm is the following simple routine, due to Kuhn [19], for computing defective edge coloring in $O(1)$ time. We describe it for the sake of completeness. Let $p'$, $1 \leq p' \leq \Delta$, be a parameter. Each vertex $v$ labels the edges $e_1, e_2, ..., e_{deg(v)}$ that are incident to $v$ with labels in $\{1, 2, ..., p'\}$, so that there is no label that is assigned to more than $\lceil \Delta/p' \rceil$ edges. (For example, it can label the edges $e_1, e_2, ..., e_{\lceil \Delta/p' \rceil}$ with 1, $e_{\lceil \Delta/p' \rceil+1}, e_{\lceil \Delta/p' \rceil+2}, ..., e_{2\cdot\lceil \Delta/p' \rceil}$ with 2, etc'.) For each edge $e = (u, w)$, both endpoints $u$ and $w$ send to each other the labels $\ell_u(e)$ and $\ell_w(e)$ that they assigned to the edge $e$. Finally, they set the colors $\varphi(e)$ of $e$ to be equal to the pair $(\ell_u(e), \ell_w(e))$, where the order is determined by the identities of $u$ and $w$. (For example, $\ell_u(e)$ appears before $\ell_w(e)$ if $Id(u) < Id(w)$, and after it otherwise.)

Obviously, the number of colors employed by $\varphi$ is at most $p'^2$. To analyze its defect consider an edge $e = (u, w)$. Suppose without loss of generality that $Id(u) < Id(w)$. For an edge $e' = (u, z)$ with $Id(u) < Id(z)$ to get the same color as $e$, $u$ should have label both $e$ and $e'$ with the same label. Hence there are at most $\lceil \Delta/p' \rceil$ incident edges $e' = (u, z)$ to $e = (u, w)$ with $Id(u) < Id(z)$ and $\varphi(e') = \varphi(e)$. For an edge $e' = (u, z)$ with $Id(u) > Id(z)$ to get the same color as $e$, the labels $\ell_u(e')$ and $\ell_w(e)$ must agree. However, $u$ labels at most $\lceil \Delta/p' \rceil$ edges with the label $\ell_w(e)$, and so there are at most $\lceil \Delta/p' \rceil$ incident edges $e' = (u, z)$ to $e = (u, w)$ with $Id(u) > Id(z)$ and $\varphi(e') = \varphi(e)$. Hence overall there are at most $2 \cdot \lceil \Delta/p' \rceil$ edges $e'$ that share with $e$ the vertex $u$ that satisffy $\varphi(e') = \varphi(e)$. By symmetrical considerations there are at most $2 \cdot \lceil \Delta/p' \rceil$ edges $e'$ that share with $e$ the vertex $w$ and satisfy $\varphi(e') = \varphi(e)$. Hence the overall defect of $\varphi$ is at most $4 \cdot \lceil \Delta/p' \rceil$.

**Corollary 5.4.** *[19] For any parameter $p'$, $1 \le p' \le \Delta$, a $4 \cdot \lceil \Delta/p' \rceil$-defective $p'^2$-edge-coloring can be computed in $O(1)$ time.*

Hence the $\lfloor \Lambda/(b \cdot p) \rfloor$-defective $O((b \cdot p)^2)$-edge-coloring that is required on line 1 of Algorithm 1 (Procedure Defective-Color) can also be computed in $O(1)$ time. (We remark that in our algorithm, $b \cdot p$ is never greater than $\Lambda/4$, and thus the factor 4 in the defect of the coloring provided by Corollary 5.4 can be essentially ignored.) Next, we analyze the edge-coloring variant of Procedure Defective-Color (Algorithm 1). In our implemenation, for each edge $e = (u, w)$, at all times both the endpoints $u$ and $w$ of $e$ will maintain the current color of $e$. In line 1 of the procedure the $\lfloor \Lambda/(b \cdot p) \rfloor$-defective $O((b \cdot p)^2)$-coloring $\varphi$ is computed using the routine that was described above. As a result of this routine (that requires $O(1)$ time), for each edge $e = (u, w)$, both $u$ and $w$ know $\varphi(e)$.

Consider the while-loop on lines 4 - 10 of Algorithm 1. For an edge $e = (u, w)$ with $\varphi(e) = i$ for some $i \in \{1, 2, ..., O((b \cdot p)^2)\}$, at some point its endpoints $u$ and $w$ recolor $e$, i.e., compute the $\psi$-color $\psi(e)$. To this end each of them needs to know the numbers $N_e(k) = |\{e' \in \Gamma(e) \mid \psi(e') = k, \varphi(e') < \varphi(e)\}|$, for every $k \in \{1, 2, ..., p\}$. ($\Gamma(e)$ is the set of the edges incident to $e$.) Define also $N_{e,u}(k) = |\{e' \ni u \mid e' \ne e, \psi(e') = k, \varphi(e') < \varphi(e)\}|$. Observe that $N_e(k) = N_{e,u}(k) + N_{e,w}(k)$. Since $u$ (respectively, $w$) can compute $N_{e,u}(1), N_{e,u}(2), ..., N_{e,u}(p)$ (resp., $N_{e,w}(1), N_{e,w}(2), ..., N_{e,w}(p)$) locally, it follows that for $u$ (resp., $w$) to be able to compute $N_e(1), N_e(2), ..., N_e(p)$ it needs to receive from $w$ (resp., $u$) the numbers $N_{e,w}(1), N_{e,w}, (2), ..., N_{e,w}(p)$ (resp., $N_{e,v}(1), N_{e,v}, (2), ..., N_{e,v}(p)$). This requires sending $p$ messages over each edge in each direction.

Hence Procedure Defective-Color invoked with parameters $G, b, p, \Lambda$ can be implemented for computing $((\Lambda/(b \cdot p) + \Lambda/p) \cdot 2 + 2)$-defective $p$-coloring $\psi$ in $O((b \cdot p)^2)$-time. (Observe that the neighborhood independence is $c = I(L(G)) = 2$, and that the $\log^* n$ term of the runnning time from Corollary 3.3 disappears because we use Kuhn's defective edge-coloring routine, which requires $O(1)$ time.) This implementation, however, requires to send $O(p)$ messages of size $O(\log n)$ each through edges of $G$. If one allows only short messages (i.e., of size $O(\log n)$), then the running time of the procedure grows to $O((b \cdot p)^2 \cdot p) = O(b^2 \cdot p^3)$. Observe that in both cases, as a result of this invocation, for every edge $e = (u, w)$, both endpoints $u$ and $w$ know the resulting $\varphi$-color $\varphi(e)$.

Next, we describe the edge-coloring variant of Procedure Legal-Color (Algorithm 2). On the bottom level of the recursion (line 2 of Algorithm 2) the algorithm computes a legal $(\Lambda + 1)$-coloring of a graph with maximum degree at most $\Lambda$. In the vertex coloring variant of the procedure, we used Lemma 2.1(2) for this, which is an algorithm from [4] or from [19] that requires $O(\Lambda) + \frac{1}{2} \log^* n$ time. Simulating this algorithm would incur large messages. Instead we use a $(2\Lambda - 1)$-edge-coloring algorithm of Panconesi and Rizzi [24]. This algorithm requires $O(\Lambda) + \log^* n$ time. In line 5 of algorithm 2 (Procedure Legal-Color) we invoke the edge-coloring variant of Procedure Defective-Color, which we described and analyzed above.

The recursive invocations of Procedure Legal-Color (line 8 of Algorithm 2) can be invoked seemlessly, because for every edge $e = (u, w)$ both $u$ and $w$ know the current $\psi$-color of $e$, and thus know to which subgraph $G_i$ the edge belongs. Other steps of Procedure Legal-Color can be performed locally, in zero time.

Next, we analyze the resulting edge-coloring variant of Procedure Legal-Color. The number of colors it employs can be bounded exactly in the same way as for the vertex-coloring variant of the procedure, and the resulting estimates are the same as in that case. The analysis of the running time changes slightly, and it depends on the size of messages that we are allowed to use.

First, consider the case that we are allowed to use $p$ short messages over each edge on every round. As we have seen, the running time of the invocation of Procedure Defective-Color with parameters $G, b, p, \Lambda$ is $O((b \cdot p)^2)$ in this case, instead of $O((b \cdot p)^2) + \frac{1}{2} \log^* n$ for the vertex coloring variant. Another change in the analysis of the running time is on the bottom level of the recursion, where we now need to use $O(\Lambda) + \log^* n = O(\lambda) + \log^* n$ time, instead of $O(\lambda) + \frac{1}{2} \log^* n$. By reproducing the analysis of the running time with these changes it is easy to verify that overall we obtain (1) an $O(\Delta)$-edge-coloring in $O(\Delta^\epsilon) + \log^* n$ time; (2) an $O(\Delta^{1+\eta})$-edge-coloring in $O(\log \Delta) + \log^* n$ time; and (3) a $\Delta^{1+o(1)}$-edge-coloring in $O((\log \Delta)^{1+\zeta}) + \log^* n$ time, where $\epsilon, \eta, \zeta > 0$ are arbitrarily small positive constants.

The size of the messages that the algorithm employs needs to allows sending $p$ numbers $N_{e,u}(1), N_{e,u}(2)$, ..., $N_{e,u}(p)$ in one message, where $e$ is an edge and $u$ is one of its endpoints. Each of these numbers is an integer between 1 and $\Delta$, and so overall one needs $O(p \log \Delta)$ bits to encode them. Also, for implementing the algorithm of [24] on the bottom level of the recursion one needs messages of size $O(\log n)$. Hence the total message size is $O(\max\{p \log \Delta, \log n\})$, where the value of $p$ depends on wheither we want to get (1), (2) or (3). In case (1) $p = O(\Delta^{\epsilon/3})$; in case (2) $p = O(1)$; in case (3) $p = O(\log^\zeta \Delta)$. Hence in case (2) the message size is just $O(\log n)$, as desired.

Next, consider the case that we are allowed only messages of size $O(\log n)$. Then, as we have seen, inside Procedure Defective-Color we need to send $p$ messages over each edge $e = (u, w)$ for both endpoints $u$ and $w$ to be able to compute $N_e(1), N_e(2), ..., N_e(p)$, and requires $O(p)$ time. The overall running time of Procedure Defective-Color in this case is, as was already shown, $O(b^2 \cdot p^3)$. This runnning time dominates the running time of each recursion level of Procedure Legal-Color, except for the bottom level. (On the bottom level the running time is $O(\lambda) + \log^* n$.) If we aim at getting an $O(\Delta)$-edge-coloring then $b = \lceil \Delta^{\epsilon/6} \rceil$, $p = \lceil \Delta^{\epsilon/3} \rceil$, $\lambda = \lceil \Delta^\epsilon \rceil$, and hence the number $r$ of recursion levels is $O(1)$. Hence the overall runnng time becomes $O(\Delta^{4\epsilon/3}) + \log^* n$ instead of $O(\Delta^\epsilon) + \log^* n$. Moreover, by scaling the constant $\epsilon$ (i.e., setting $\epsilon' = 3\epsilon/4$), we can get the same running time $O(\Delta^\epsilon) + \log^* n$ as in the vertex-coloring variant of the algorithm. (Except for the constant hidden by the $O$-notation in the number of colors $O(\Delta)$, which grows by a factor of 4/3.) By similar considerations we also get $\Delta^{1+o(1)}$-edge-coloring in $O(\log^{1+\zeta} \Delta) + \log^* n$ time, with slightly larger hidden constants. To summarize:

**Theorem 5.5.** *For any graph $G$ and constants $\epsilon, \eta, \zeta > 0$,*
*(1) an $O(\Delta)$-edge-coloring of $G$ can be computed in $O(\Delta^\epsilon) + \log^* n$ time.*
*(2) an $O(\Delta^{1+\eta})$-edge-coloring of $G$ can be computed in $O(\log \Delta) + \log^* n$ time.*
*(3) a $\Delta^{1+o(1)}$-edge-coloring of $G$ can be computed in $O((\log \Delta)^{1+\zeta}) + \log^* n$ time.*
*Moreover, the algorithm that computes these colorings employs messages of size $O(\log n)$.*

The constants hidden by the notation here are somewhat larger than the corresponding constants in Theorem 4.8. One can also keep these constants as small as there, at the expanse of increasing the message size to $O(\max\{\Delta^{\epsilon/3}, \log n\})$ in (1), $c' \cdot \log n$ for a very large constant $c'$ in (2), and $O(\max\{\log^{1+\zeta} \Delta, \log n\})$ in (3).

# 6 Extensions

In this section we present two simple extensions of our main results (that are given by Theorems 4.8 and 5.3). In Section 6.1 we combine our new deterministic algorithm with an existing randomized routine of [20]. In this way we derive a randomized $O(\Delta \cdot \min\{\Delta, \log n\}^\epsilon)$-edge-coloring algorithm with running time $O(\log \log n)$, for any arbitrarily small constant $\epsilon > 0$. For $\Delta \leq \log^{1-\delta} n$, for some fixed constant $\delta > 0$, this result compares favorably with all previous results. Specifically, the variant of the randomized algorithm of [29] that requires $O(\log \log n)$ time employs $\Omega(\log n)$ colors, even when $\Delta$ is small.

In Section 6.2 we extend our deterministic tradeoff, and obtain a much faster algorithm that uses much more colors. Specifically, this variant of our algorithm computes an $O(\frac{\Delta^2}{g(\Delta)})$-edge-coloring of a general input graph in just $O(\log g(\Delta)) + \log^* n$ time, for any monotonic non-decreasing function $g()$.

## 6.1 A randomized algorithm

In this section we combine our algorithm with a randomized defective coloring routine, due to Kuhn and Wattenhofer [20]. As a result we obtain a randomized algorithm that works for graphs of bounded neighborhood independence. For any constant $\eta > 0$, it produces an $O(\Delta \cdot \min\{\Delta, \log n\}^\eta)$-vertex-coloring in $O(\log \log n \cdot \frac{\log^* n}{\log(\log^* n)})$ time, with high probability.

If $\Delta = O(\log n)$ the or algorithm from Theorem 4.8(2) provides the desired result. Hence in the sequel we assume that $\Delta = \omega(\log n)$.

We start with describing the routine from [20] that produces an $O(\log n)$-defective $O(\frac{\Delta}{\log n})$-vertex-coloring of general graphs locally. This routine and its analysis are provided here for the sake of completeness. Each vertex $v$ picks a color from the set $\{1, 2, ..., \lceil \Delta / \log n \rceil\}$ independently uniformly at random. The expected defect of $v$, i.e., the number of neighbors of $v$ that get the same color as $v$ does, is at most $\log n$. Denote the defect of $v$ by $defect(v)$. Hence, by Chernoff's bound, for any constant $\kappa > 1$,

$$P(defect(v) > \kappa \cdot e \cdot \ln n) < \left(\frac{e^{\kappa-1}}{(\kappa e)^{\kappa e}}\right)^{\ln n} \leq \frac{1}{n^{(\ln \kappa)\kappa e+1}}.$$

Hence with probability at least $\frac{1}{n^{(\ln \kappa)\kappa e}}$, the defect of the resulting coloring is $O(\kappa \cdot \log n)$. Once this defective coloring $\varphi$ is computed we apply our algorithm (from Theorem 4.8(2)) on each of the $\varphi$-color classes $G_1, G_2, ..., G_{\lceil \Delta / \log n \rceil}$. Each of these subgraphs has bounded neighborhood independence, and maximum degree at most $O(\kappa \cdot \log n) = O(\log n)$. (As $\kappa = O(1)$.) Hence in $O(\log \log n \cdot \frac{\log^* n}{\log(\log^* n)})$ time our algorithm produces a $(\log n)^{1+\eta}$-vertex-coloring for each of these subgraphs. Hence it is an $O(\Delta \cdot (\log n)^\eta)$-vertex-coloring of the original graph $G$.

To summarize:

**Theorem 6.1.** *For an arbitrarily small constant $\eta > 0$, an arbitrarily large constant $\kappa$, and a graph $G$ of bounded neighborhood independence, our algorithm constructs an $O(\Delta \cdot \min\{\Delta, \log n\}^\eta)$-vertex-coloring of $G$ in time $O(\log \log n \cdot \frac{\log^* n}{\log(\log^* n)})$, with probability at least $1 - \frac{1}{n^\kappa}$.*

Similarly to our other results, this result too can be slightly specialized (and slightly improved) to the case of edge-coloring of general graphs. In this case an $O(\log n)$-defective $O(\frac{\Delta}{\log n})$-edge-coloring routine requires $O(1)$ rounds, because we need that for every edge $e = (u, w)$, both its endpoints will know the computed color. Once the defective edge coloring is computed, we invoke our algorithm from Theorem 5.5(2) on each of the (edge-disjoint) subgraphs. The algorithm requires just $O(\log \log n)$ time, and uses the same number of colors as stated in Theorem 6.1.

**Corollary 6.2.** *For an arbitrarily small constant $\eta > 0$, and an arbitrarily large constant $\kappa$, our algorithm $O(\Delta \cdot \min\{\Delta, \log n\}^\eta)$-edge-colors general graphs, in $O(\log \log n)$ time, with probability at least $1 - \frac{1}{n^\kappa}$.*

## 6.2 A tradeoff

Next, we argue that for any monotonic non-decreasing function $g(\cdot)$, one can get an $O(\frac{\Delta^2}{g(\Delta)})$-vertex-coloring of graphs with bounded neighborhood independence in $O(\log g(\Delta) \cdot \frac{\log^* g(\Delta)}{\log(\log^* g(\Delta))}) + \frac{1}{2} \log^* n$ time. This result also translates into an $O(\frac{\Delta^2}{g(\Delta)})$-edge-coloring algorithm with running time $O(\log g(\Delta)) + \log^* n$. This extension is closely to the vertex-coloring tradeoff presented in [5]. Specifically, there it is shown that one can produce an $O(\frac{\Delta^2}{g(\Delta)})$-vertex-coloring of a general graph in $O(\log n \cdot \log g(\Delta))$ time.

For a small positive constant $0 < \eta < 1$, set $q(\Delta) = g(\Delta)^{\frac{1}{1-\eta}}$, and $p = p(\Delta) = \frac{\Delta}{q(\Delta)}$. Compute a $\frac{\Delta}{p}$-defective $O(p^2)$-coloring $\varphi$ of $G$ in time $O(\log^* n)$ using Lemma 2.1(3) (an algorithm from [19]). This coloring partitions $G$ into $O(p^2)$ vertex disjoint subgraphs $G_1, G_2, ..., G_{O(p^2)}$, each of maximum degree at most $\frac{\Delta}{p}$. Since the original graph $G$ has neighborhood independence at most $c$, by Lemma 3.6 it follows that each of the subgraphs $G_1, G_2, ..., G_{O(p^2)}$ has neighborhood independence at most $c$ as well.

On each of these subgraphs in parallel, invoke our algorithm whose properties are summarized in Theorem 4.8(2). Within $O(\log \frac{\Delta}{p} \cdot \frac{\log^*(\Delta/p)}{\log(\log^*(\Delta/p))}) + \frac{1}{2} \log^* n$ time we obtain an $O((\frac{\Delta}{p})^{1+\eta})$-vertex-coloring of each of these subgraphs. Overall we obtain an $O(p^2 \cdot (\frac{\Delta}{p})^{1+\eta}) = O(p^{1-\eta} \cdot \Delta^{1+\eta})$-vertex-coloring of $G$ within $O(\log \frac{\Delta}{p} \cdot \frac{\log^*(\Delta/p)}{\log(\log^*(\Delta/p))} + \log^* n)$ time. Since $p = p(\Delta) = \frac{\Delta}{q(\Delta)}$, we have an $O(\frac{\Delta^2}{q(\Delta)^{1-\eta}})$-vertex-coloring within $O(\log q(\Delta) \cdot \frac{\log^* q(\Delta)}{\log(\log^* q(\Delta))} + \log^* n)$ time. Finally, since $g(\Delta) = q(\Delta)^{1-\eta}$, we derive an $O(\frac{\Delta^2}{g(\Delta)})$-vertex-coloring within $O(\log g(\Delta) \cdot \frac{\log^* g(\Delta)}{\log(\log^* g(\Delta))} + \log^* n)$ time.

One can also compute an auxiliary $O(\Delta^2)$-coloring $\rho$ in the beginning of the computation in $\frac{1}{2} \log^* n$ time using the algorithm of Szegedy and Vishwanathan [27]. This coloring can be used to compute a $\frac{\Delta}{p}$-defective $O(p^2)$-coloring $\varphi$ in just $O(\log^* \Delta)$ time. As we have seen in Section 3 it can also be used to eliminate the $\frac{1}{2} \log^* n$ term from the running time of the algorithm from Theorem 4.8. To summarize:

**Corollary 6.3.** *For any monotonic non-decreasing function $g(\cdot)$, one can compute an $O(\frac{\Delta^2}{g(\Delta)})$-vertex-coloring of a graph with bounded neighborhood independence in $O(\log g(\Delta) \cdot \frac{\log^* g(\Delta)}{\log(\log^* g(\Delta))}) + \frac{1}{2} \log^* n$ time.*

By the techniques that are described in Section 5 we can also $O(\frac{\Delta^2}{g(\Delta)})$-edge-color general graphs in $O(\log g(\Delta) \cdot \frac{\log^* g(\Delta)}{\log(\log^* g(\Delta))}) + \frac{1}{2} \log^* n$ time, and also in $O(\log g(\Delta)) + \log^* n$ time. The latter running time is achieved using only messages of size $O(\log n)$.

## Acknowledgements

The authors thank Alessandro Panconesi for sending them several papers and addressing their questions.

## References

[1] B. Awerbuch, A. V. Goldberg, M. Luby, and S. Plotkin. Network decomposition and locality in distributed computation. In *Proc. of the 30th Symposium on Foundations of Computer Science*, pages 364–369, 1989.

[2] J. Andrews, and M. Jacobson. On a generalization of a chromatic number. *Congressus Numer*, 47:33-48, 1985.

[3] L. Barenboim, and M. Elkin. Sublogarithmic distributed MIS algorithm for sparse graphs using Nash-Williams decomposition. In *Proc. of the 27th ACM Symp. on Principles of Distributed Computing*, pages 25–34, 2008.

[4] L. Barenboim, and M. Elkin. Distributed ($\Delta + 1$)-coloring in linear (in $\Delta$) time. In *Proc. of the 41th ACM Symp. on Theory of Computing*, pages 111-120, 2009.

[5] L. Barenboim, and M. Elkin. Deterministic distributed vertex coloring in polylogarithmic time. To appear in *Proc. of the 29th ACM Symp. on Principles of Distributed Computing*, 2010.

[6] M. Chudnovsky, and P. Seymour. The structure of claw-free graphs. *Surveys in Combinatorics 2005, London Math Soc. Lecture Note Series*, 327:153-171, 2005.

[7] L. Cowen, R. Cowen, and D. Woodall. Defective colorings of graphs in surfaces: partitions into subgraphs of bounded valence. *Journal of Grah Theory*, 10:187–195, 1986.

[8] L. Cowen, W. Goddard, and C. Jesurum. Coloring with defect. In *Proc. of the 8th ACM-SIAM Symp. on Discrete Algorithms, New Orleans, Louisiana, USA*, pages 548–557, January 1997.

[9] A. Czygrinow, M. Hanckowiak, and M. Karonski. Distributed $O(\Delta \log n)$-edge-coloring algorithm. In *Proc. of the 9th Annual European Symposium on Algorithms*, pages 345-355, 2001.

[10] D. Dubhashi, D. Grable, and A. Panconesi. Nearly-optimal, distributed edge-colouring via the nibble method. *Theoretical Computer Science*, 203:225–251, 1998.

[11] D. Durand, R. Jain, and D. Tseytlin. Applying randomized edge coloring algorithms to distributed communication: an experimental study. In *Proc. of the 7th Annual ACM Symposium on Parallel Algorithms and Architectures*, pages 264–274, 1995.

[12] M. Frick. A survey of $(m, k)$-colorings. *Quo Vadis, Graph Theory?*, volume 55 of Annals of Discrete Mathematics, pages 45-58, 1993.

[13] B. Gfeller, and E. Vicari. A randomized distributed algorithm for the maximal independent set problem in growth-bounded graphs. In *Proc. of the 26th ACM Symp. on Principles of Distributed Computing*, pages 53–60, 2007.

[14] D. Grable, and A. Panconesi. Nearly optimal distributed edge coloring in $O(\log \log n)$ rounds *Random Structures and Algorithms*, 10(3):385–405, 1998.

[15] F. Harary, and K. Jones. Conditional colorability II: Bipartite variations. *Congressus Numer*, 50:205-218, 1985.

[16] R. Jain, K. Somalwar, J. Werth, and J. C. Browne. Scheduling parallel I/O operations in multiple bus systems. *ELSEVIER Journal of Parallel and Distributed Computing*, 16(4):352-362, 1992.

[17] F. Kuhn, T. Moscibroda, and R. Wattenhofer. On the Locality of Bounded Growth. In*Proc. of the 24rd ACM Symp. on Principles of Distributed Computing*, pages 60–68, 2005.

[18] K. Kothapalli, C. Scheideler, M. Onus, and C. Schindelhauer. Distributed coloring in $O(\sqrt{\log n})$ bit rounds. In *Proc. of the 20th International Parallel and Distributed Processing Symposium*, 2006.

[19] F. Kuhn. Weak graph colorings: distributed algorithms and applications. In *proc. of the 21st ACM Symposium on Parallel Algorithms and Architectures*, pages 138–144, 2009.

[20] F. Kuhn, and R. Wattenhofer. On the complexity of distributed graph coloring. In *Proc. of the 25th ACM Symp. on Principles of Distributed Computing*, pages 7–15, 2006.

[21] N. Linial. Locality in distributed graph algorithms. *SIAM Journal on Computing*, 21(1):193–201, 1992.

[22] N. Linial and M. Saks. Low diameter graph decomposition. *Combinatorica* 13: 441 - 454, 1993.

[23] G. J. Minty. On maximal independent sets of vertices in claw-free graphs. *Journal of Combinatorial Theory*, Series B 28 (3): 284-304, 1980.

[24] A. Panconesi, and R. Rizzi. Some simple distributed algorithms for sparse networks. *Distributed computing*, 14(2):97–100, 2001.

[25] A. Panconesi, and A. Srinivasan. On the complexity of distributed network decomposition. *Journal of Algorithms*, 20(2):581-592, 1995.

[26] A. Panconesi, and A. Srinivasan. Distributed edge coloring via an extension of the Chernoff-Hoeffding bounds. *SIAM Journal on Computing*, 26(2):350–368, 1997.

[27] M. Szegedy, and S. Vishwanathan. Locality based graph coloring. In *Proc. 25th ACM Symposium on Theory of Computing*, pages 201-207, 1993.

[28] J. Schneider, and R. Wattenhofer. A log-star distributed Maximal Independent Set algorithm for Growth Bounded Graphs. In *Proc. of the 27th ACM Symp. on Principles of Distributed Computing*, pages 35–44, 2008.

[29] J. Schneider, and R. Wattenhofer. A new technique for distributed symmetry breaking. To appear in *Proc. of the 29th ACM Symp. on Principles of Distributed Computing*, 2010.

[30] V. G. Vizing. On an estimate of the chromatic class of a p-graph. *Diskret Analiz*, 3:25–30, 1964.